\documentclass{article}

\usepackage{PRIMEarxiv}

\usepackage[utf8]{inputenc} % allow utf-8 input
\usepackage[T1]{fontenc}    % use 8-bit T1 fonts
\usepackage{hyperref}       % hyperlinks
\usepackage{url}            % simple URL typesetting
\usepackage{booktabs}       % professional-quality tables
\usepackage{amsfonts}       % blackboard math symbols
\usepackage{nicefrac}       % compact symbols for 1/2, etc.
\usepackage{microtype}      % microtypography
\usepackage{lipsum}
\usepackage{fancyhdr}       % header
\usepackage{graphicx}       % graphics
\graphicspath{{media/}}     % organize your images and other figures under media/ folder
\usepackage{wrapfig}

\title{\textnormal{The Role of ChatGPT in Democratizing Data Science: An Exploration of AI-facilitated Data Analysis in Telematics}}
\author{
  Ryan Lingo \\
  Honda Research Institute, USA\\
  \texttt{ryan\_lingo@honda-ri.com} \\
}

\begin{document}
\maketitle

\begin{abstract}
The realm of data science, once reserved for specialists, is undergoing a revolution with the rapid emergence of generative AI, particularly through tools like ChatGPT. This paper posits ChatGPT as a pivotal bridge, drastically lowering the steep learning curve traditionally associated with complex data analysis. By generating intuitive data narratives and offering real-time assistance, ChatGPT democratizes the field, enabling a wider audience to glean insights from intricate datasets. A notable illustration of this transformative potential is provided through the examination of a synthetically generated telematics dataset, wherein ChatGPT aids in distilling complex patterns and insights. However, the journey to democratization is not without its hurdles. The paper delves into challenges presented by such AI, from potential biases in analysis to ChatGPT's limited reasoning capabilities. While the promise of a democratized data science landscape beckons, it is imperative to approach this transition with caution, cognizance, and an ever-evolving understanding of the tool's capabilities and constraints.
\end{abstract}

% keywords can be removed

\section{Introduction}
Large language models, such as ChatGPT developed by OpenAI, are revolutionizing multiple sectors within society, including data science. ChatGPT, an autoregressive language model trained on diverse internet text, excels at interpreting and generating human-like text. This capability introduces a novel approach to data science, offering immense potential in multiple stages of data analysis, such as data cleaning, exploratory data analysis, visualization, and the explanation of complex statistical concepts.

For beginners and non-technical domain experts, ChatGPT can become a revolutionary tool. It provides an accessible interface, allowing users to interact in natural language, thus reducing the struggle with complex programming syntax or statistical jargon. This feature enables a sharper focus on understanding and interpreting data, the essence of data analysis, democratizing the field of data science by lowering the barriers to entry.

Data science often relies heavily on high-level, general-purpose programming languages like Python, which offers simplicity and a host of powerful libraries designed for data analysis. Among these libraries, Pandas stands out as an open-source tool for data manipulation and analysis, providing flexible data structures for fast and efficient work with structured data.

In this paper, we will demonstrate the use of Python and Pandas in our data analysis examples, showing how ChatGPT can provide valuable assistance. Notably, ChatGPT demonstrates a robust understanding of Python and Pandas; it can interpret and generate code, proving itself to be an effective tool for both coding assistance and learning. By offering a natural language interface to Python and Pandas, ChatGPT is poised to lower the learning curve for beginners and non-tech domain experts, further emphasizing its role in democratizing data science.

\subsection{Purpose and Organization of the Paper}
The primary aim of this paper is to explore and elucidate how ChatGPT, a large language model, can aid in key areas of data analysis, specifically data cleaning, feature engineering, exploratory data analysis, and data visualization. By focusing on these fundamental stages of data analysis, we intend to show how ChatGPT can contribute to democratizing the field of data science, making it more accessible to beginners and domain experts from non-technical fields.
This paper is organized as follows:
\begin{itemize}
\item \textbf{Background:} We begin by offering background on natural language processing, the rise of large language models, and a brief literature review.
\item \textbf{Applications of ChatGPT in Data Analysis:} We then dive into specific use-cases of using ChatGPT for data analysis. We concentrate on tasks such as cleaning and preprocessing data, generating and engineering new features from raw telematics data, performing exploratory data analysis, and creating effective data visualizations.
\item \textbf{Democratization of Data Science via ChatGPT:} Subsequently, we discuss the broader impact of ChatGPT and similar models in democratizing data science. This includes ways it lowers the barriers to entry and how it enables faster iterations in the data analysis process.
\item \textbf{Limitations and Challenges:} The paper also addresses potential limitations and challenges in using ChatGPT. These concerns range from the risk of bias in the AI model to limitations in its reasoning capability and the need for human oversight and verification.
\item \textbf{Conclusions and Future Directions: } Finally, we conclude with a summary of how ChatGPT can aid in data analysis and discuss the need for responsible development and use of AI systems. We also touch upon potential future opportunities for enhancing data science through human-AI collaboration.
\end{itemize}

Our exploration of the capabilities, implications, and limitations of ChatGPT in data science aims to contribute to the ongoing discourse on the role of AI in democratizing technical domains. This focused investigation into specific tasks within data analysis serves to showcase the practical benefits and potential challenges of integrating AI tools into data science workflows.

\subsection{Potential Benefits and Challenges of Using ChatGPT in Data Analysis}
Large language models like ChatGPT offer several significant advantages that can streamline and enhance data analysis. By understanding and generating code, ChatGPT makes it possible to perform complex data analysis tasks using natural language. This feature not only simplifies the data analysis process but also makes it more accessible, particularly for those who may not have extensive coding expertise. Furthermore, ChatGPT's ability to handle diverse tasks --- from data cleaning, feature engineering, exploratory data analysis to data visualization --- makes it an invaluable tool across the data analysis pipeline.

On the other hand, the use of ChatGPT is not without challenges. One notable concern is the potential for biases in the training data to be reflected in the model's outputs. These biases could inadvertently skew analysis or lead to erroneous conclusions if not appropriately accounted for. Additionally, despite their impressive capabilities, large language models like ChatGPT lack a true understanding or reasoning ability beyond their training data. This limitation means they may sometimes produce outputs that are plausible-sounding but factually incorrect or nonsensical. Finally, there is a risk of overreliance on ChatGPT at the expense of human understanding and intuition, which are critical in data analysis. Therefore, it is important to use these models responsibly and always verify the outputs they generate.

This paper will delve into these benefits and challenges in greater depth, providing a nuanced view of the use of ChatGPT in data analysis. Our exploration aims to shed light on the practical implications of AI integration into data science and provide considerations for responsible usage to ensure a balanced human-AI collaboration.

% \section{Headings: first level}
\section{Background}
\subsection{A Brief History of Natural Language Processing and The Rise of Large Language Models}
% \label{sec:headings}
Natural Language Processing (NLP), a unique branch of artificial intelligence that emerged in the 1950s, was developed to enable computers to understand and work with human language. The early approaches in this field predominantly centered on rule-based methods. These systems, reliant on manually curated rules and lexicons, were quite rigid and had significant limitations in terms of scalability and maintenance.

The landscape of NLP began to change in the 1990s with the advent of statistical methods. This new paradigm utilized probabilistic models and machine learning algorithms, allowing NLP systems to learn from data and thereby perform more robustly across a wider range of tasks.

In the past decade, deep learning has radically transformed NLP. Innovations like recurrent neural networks (RNNs), long short-term memory (LSTM) units, and attention mechanisms have facilitated the processing of sequential data and the capturing of long-range dependencies in text, thus significantly enhancing the performance of tasks such as machine translation, named entity recognition, and text generation.

The real game-changer, however, came with the introduction of the transformer architecture in 2017. Presented in the paper titled "Attention is All You Need" by Vaswani et al. [6], transformers use self-attention mechanisms to weigh the importance of different words in a sentence, resulting in greatly improved performance on a myriad of tasks. This architecture forms the backbone of the current generation of large language models, including OpenAI's GPT series.

Large language models like GPT-2 and GPT-3 have taken NLP to unprecedented heights. Trained on vast amounts of text data, these models can generate impressively human-like text, answer questions, translate languages, and even understand and generate code.

ChatGPT, a variant of the GPT series developed by OpenAI, is particularly tuned for conversation-like interaction. It is this unique model that we will be examining in this paper for its potential to democratize data science by assisting in various stages of data analysis.
\subsection{Literature Review}
In the rapidly evolving field of data science, large language models (LLMs), particularly ChatGPT, have emerged as potentially transformative tools. Awan [1] provides an overarching, albeit general, framework for integrating ChatGPT into data science workflows. However, it falls short of offering explicit guidance for its application within the mobility and telematics sector. 

Lingo [2] expands upon this foundation with a focused exploration into the generation of synthetic datasets through ChatGPT in the telematics realm. His work underscores the model's potential to enhance data availability, opening doors for innovative research opportunities in complex sectors.

Hu et al. [3] push the boundaries of ChatGPT's applications by successfully employing the model for drug trafficking detection on social media, demonstrating its ability to resolve complex issues when given knowledge-informed prompts. This study accentuates the potential of LLMs like ChatGPT in enhancing data accessibility and availability.

Despite the compelling prospects of LLMs, Scheider et al. [4] caution against their unbridled use in the education sector, particularly in Geography and GIScience, citing potential threats to the validity of assessments. This critical evaluation invites contemplation of the potential challenges and limitations of employing LLMs like ChatGPT in educational contexts.

Complementing this discourse, Tu et al. [5] scrutinize the potential impacts of LLMs on data science education, shedding light on their potential benefits as well as ethical and privacy concerns. Although the study provides a general overview, its implications are pertinent to the specific application of LLMs to telematics data, which forms the crux of this paper.

Collectively, these studies enrich the understanding of ChatGPT's potential in data science, fostering an inclusive debate on its benefits and limitations. However, while these studies discuss various facets of ChatGPT, they do not attempt to argue how ChatGPT is democratizing data science, which is the central thesis of this paper. Moreover, there remains a clear gap in the literature concerning the model's application within the mobility and telematics sector, which this paper aims to address.

\section{Applications of ChatGPT in Data Analysis}
\subsection{Introduction}
The complexity and volume of telematics data often make its analysis a formidable task. The application of AI, in particular natural language processing models like ChatGPT, can substantially streamline and expedite this process. This section of the paper is devoted to exploring the role of ChatGPT in the analysis of telematics data, specifically its utility in generating a synthetic dataset, crafting summary statistics, creating features, conducting exploratory data analysis, and designing data visualizations.

Our journey into telematics data analysis with ChatGPT begins with the generation of a synthetic dataset. This dataset, crafted by the model, will be used throughout the paper to illustrate the various stages of data analysis and the ways ChatGPT can contribute to each.

Following dataset generation, we delve into the fundamental stage of understanding the data in hand. ChatGPT aids in generating summary statistics of telematics data, thereby illuminating the underlying structure, variability, and potential issues such as missing values in the data.

Next, the emphasis will be on the transformation of raw and unprocessed telematics data into a form more suitable for analysis or modeling. This involves the creation of new features, a process where ChatGPT's capabilities will be highlighted.

Subsequently, the focus shifts towards exploratory data analysis (EDA). An essential step in data analysis, EDA facilitates the identification of patterns, the spotting of anomalies, hypothesis testing, and assumption checking. The usefulness of ChatGPT in enhancing EDA by offering insights and interpretations in natural language will be explored.

Through this thorough exploration of core aspects of data analysis, we aim to provide a detailed perspective on how AI, specifically ChatGPT, can be leveraged for telematics data analysis.

\subsection{Telematics Data: an Introduction}
Telematics data fundamentally refers to the digital records produced by the telematics systems onboard a vehicle. These systems, using an array of sensors and GPS technology embedded within the vehicle, amass diverse facets of vehicular behavior. This includes parameters like location, speed, acceleration, deceleration, and fuel usage, as well as diagnostic information potentially indicating engine problems. 

The richness and granularity of telematics data make it an invaluable resource for many sectors. For example, those in charge of managing fleets can utilize this data as a tool for overseeing vehicle conditions, optimizing operational expenses, and boosting the total efficiency of their fleet. Alternatively, insurance firms can employ this data to formulate policies based on usage, where premium amounts are determined according to the specific driving habits of an individual.

Despite its value, the analysis of telematics data is challenging. The data is inherently high-dimensional, often involving numerous variables recorded over time. It is also often unstructured and may contain anomalies, making preprocessing a crucial step in the analysis pipeline. Furthermore, the analysis techniques required can be complex, involving a blend of time-series analysis, machine learning, and, in some cases, deep learning methods.

In the following sections, we will demonstrate how ChatGPT can assist with each stage of the analysis, from data cleaning and feature engineering to exploratory analysis and visualization.

\subsection{Creation of Synthetic Telematics Data}
As emphasized in my previous research [2], synthetic datasets, while inherently artificial, hold significant potential in a myriad of research scenarios. One of the pivotal advantages of these datasets lies in their relevance when real-world data accessibility becomes a bottleneck, often due to issues ranging from privacy concerns to sheer unavailability. In the ever-evolving world of data-driven research, where real datasets can often be riddled with issues such as missing data, inconsistencies, or potential bias, synthetic datasets emerge as a beacon of reliability and uniformity.

For the context of this paper, it's essential to note that the generation of a synthetic telematics dataset transcends the confines of a typical academic exercise. This isn't about generating data for the sake of it. Instead, it serves a broader, more instrumental purpose: providing a concrete, meticulously structured dataset that becomes the backbone of our exploration. Through this lens, we aim to shed light on the multifaceted capabilities of ChatGPT, especially in the realm of data analysis. By anchoring our explorations to this dataset, we're not just ensuring a consistent reference point for our examples, but we're also fortifying the clarity and comprehensibility of our demonstrations. We're setting a stage where every move, every transformation, is traceable and resonates with the reader.

But why, one might wonder, lean on a synthetic dataset in such a context? There are a multitude of reasons, some of which are:
\begin{enumerate}
  \item \textbf{Demonstrative Consistency:}  A singular, cohesive synthetic dataset forms the linchpin of our entire exploration. It's this consistent reference that ensures that readers, irrespective of their background, can seamlessly follow along, grasping the nuances of the transformations and deriving insights at every pivotal juncture.

  \item \textbf{Privacy and Ethical Assurance:} In an age where data privacy is paramount, the impetus is on researchers to tread cautiously. Leveraging synthetic data serves as our conscious step in this direction, ensuring we're not wading into murky waters by unintentionally revealing sensitive or personal data. This commitment ensures our alignment with the gold standards of ethical research.

   \item \textbf{Tailored Complexity:} Every research has its unique demands, its unique set of challenges. By taking the reins and crafting our own dataset, we're not just passively accepting data. We're actively molding it, introducing intricacies, complexities, or even anomalies that are bespoke to our requirements. This allows us to vividly highlight the diverse capabilities of ChatGPT, be it in data cleaning, intricate analysis, or even the visualization of patterns hitherto unseen.

   \item \textbf{Accessibility:} Using a synthetic dataset ensures that any researcher, student, or enthusiast can engage with our work without the hindrance of sourcing proprietary or hard-to-find datasets. This promotes a more inclusive scientific environment where knowledge is more broadly disseminated, allowing individuals from varied backgrounds and resources to participate in the discourse.

\end{enumerate}

In the grand tapestry of data-driven research, synthetic datasets aren't just threads; they're the vibrant colors that make the entire picture come alive. As we delve deeper into this paper, we'll unravel the ways in which this dataset, a product of precision and forethought, plays a pivotal role in our journey with ChatGPT.

To achieve this, we posed the following prompt to ChatGPT:

\begin{figure}[h!]
\begin{center}
\includegraphics[width=165mm]{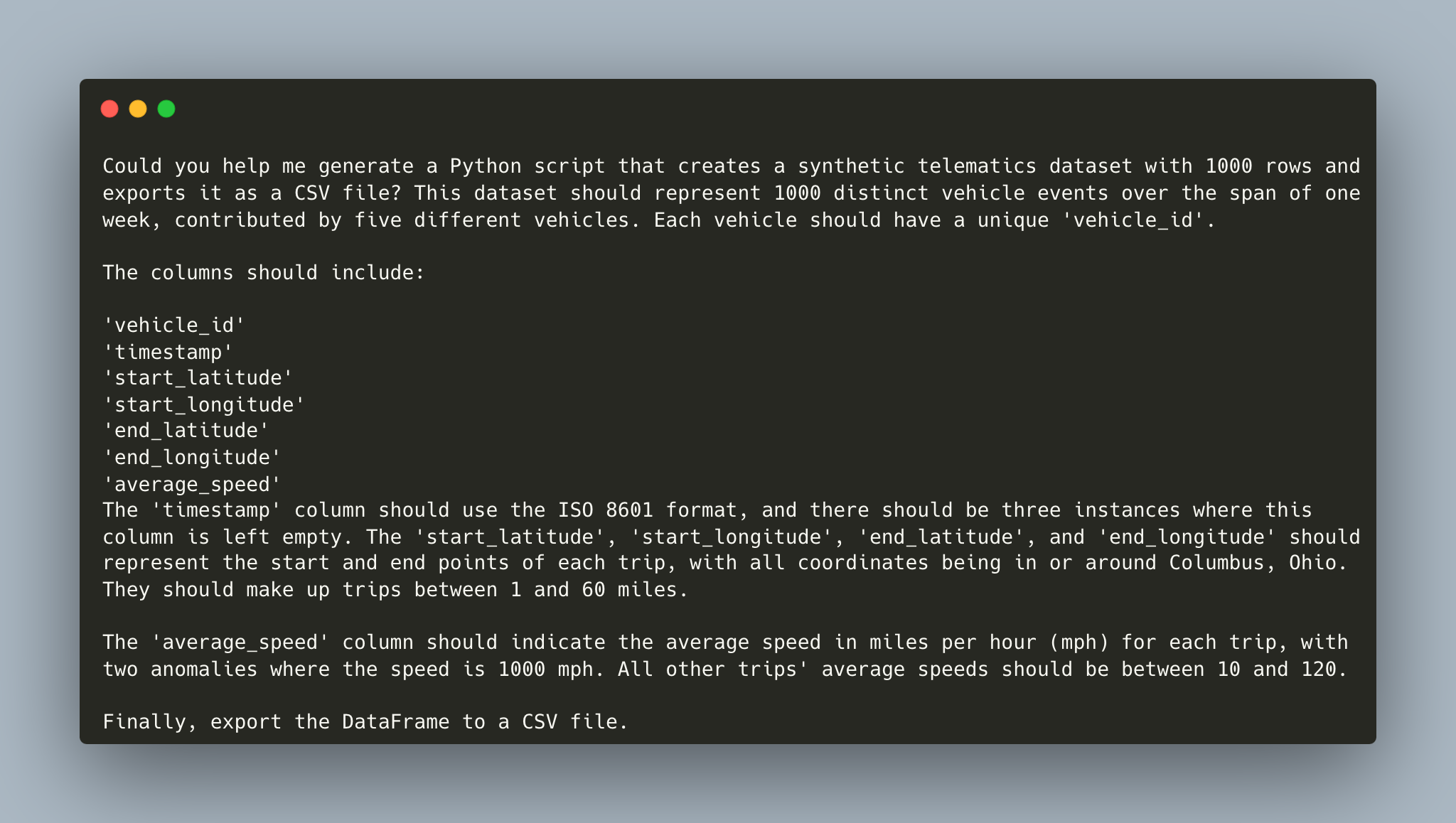}
\caption{Initial prompt to generate synthetic telematics data.}
\end{center}
\end{figure}

With this perspective in mind, let's detail the Python script that leverages ChatGPT's capabilities to generate this synthetic dataset. This dataset will serve as the cornerstone for the subsequent sections of this paper. However, a word of caution: due to the inherent variability and probabilistic nature of generative models like ChatGPT, the dataset you generate may not be identical to the one I produced. This is because ChatGPT, when given the same prompt, can yield diverse outputs based on its internal mechanics and stochasticity.
\begin{figure}[h!]
\begin{center}
\includegraphics[width=150mm]{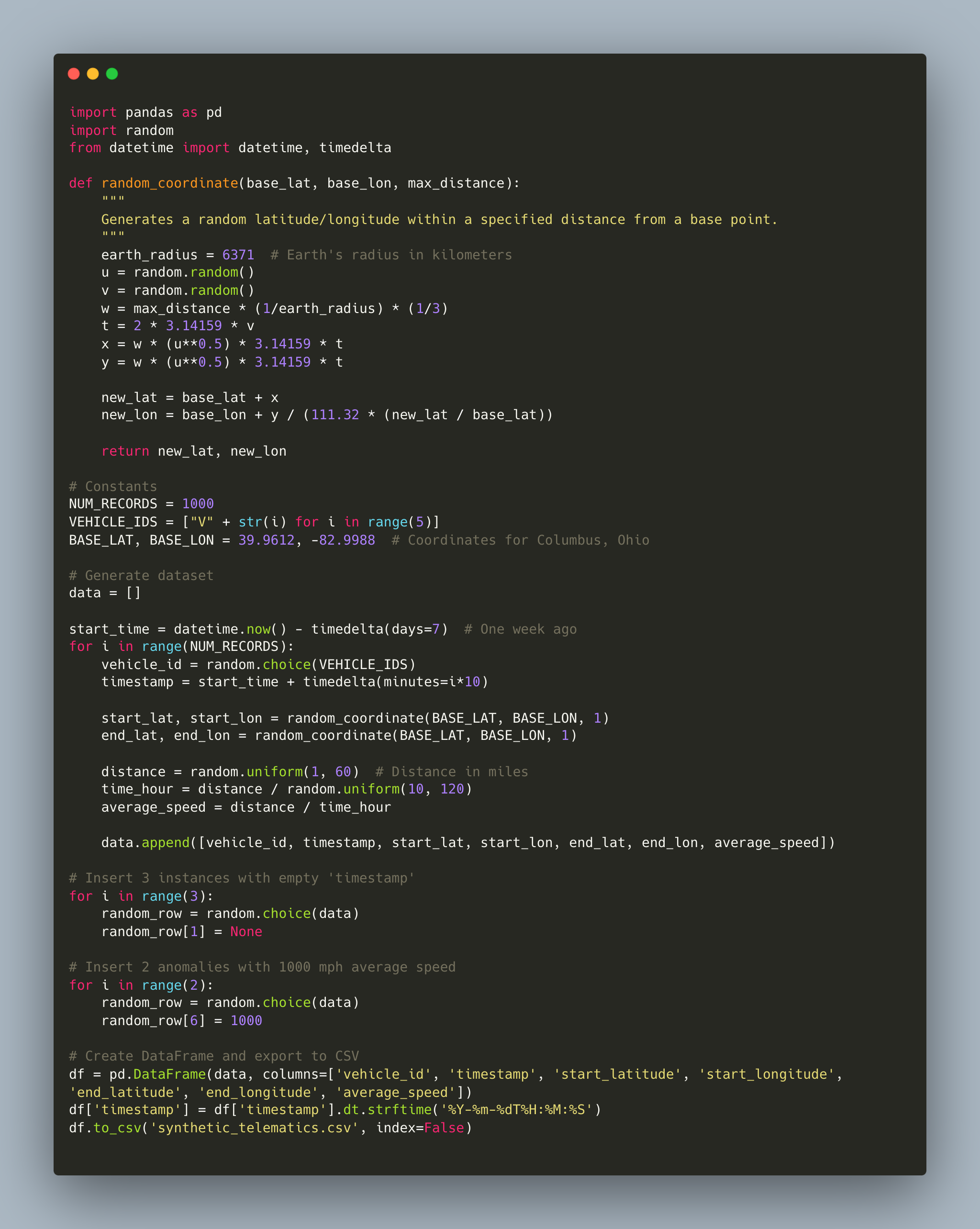}
\caption{ChatGPT output code to generate data.}
\end{center}
\end{figure}

% \begin{figure}[h]
% \begin{center}
% \includegraphics[width=165mm]{code_to_generate_dataset.png}
% \caption{Initial prompt to generate synthetic telematics data.}
% \end{center}
% \end{figure}

Now, we’ll utilize this Python script to generate the dataset. As depicted in Figure 2, you can find the code output provided by ChatGPT. Before proceeding, if you are working in a fresh environment, it may be necessary to install some requisite libraries to ensure smooth execution. This dataset will then serve as the foundation for our subsequent analyses in this paper.

\subsection{Data Cleaning and Preprocessing}
\subsubsection{Introduction}
In the intricate domain of data science, raw telematics data rarely presents itself in an ideal state. With diverse sources, immense granularity, and real-time generation, the need for preprocessing becomes unequivocally paramount. This stage metamorphoses raw, noisy information into structured and coherent data, laying the foundation for consequential insights and analyses.

\subsubsection{Significance of Data Cleaning in Telematics:}

\begin{itemize}
\item \textbf{Handling Missing Values: } Given the real-time generation of telematics data, gaps are inevitable. Occasional missing timestamps or GPS discrepancies can lead to fragmented data entries that hinder comprehensive analysis.
\item \textbf{Managing Outliers:} Sensor anomalies or transient glitches can introduce outliers, such as a speed reading of 1000 mph. Detecting and rectifying these irregularities is essential for maintaining the integrity of the dataset.
\item \textbf{Ensuring Data Integrity:} With data pouring in from various vehicles, sensors, and onboard systems, maintaining a consistent standard across the dataset is non-negotiable.
\item \textbf{Enhancing Data Quality:} The caliber of insights—whether visual representations of driving behaviors or predictive analytics—is directly proportional to the quality of the input data.
\end{itemize}
While the significance of data cleaning is universal in data analytics, its imperative nature in telematics is accentuated. Given the vast volumes and real-time implications, any misinterpretation or oversight can culminate in flawed strategies, inaccurate vehicle maintenance predictions, or even safety hazards.

However, it's essential to highlight that our synthetic dataset, crafted specifically for this analysis, isn't devoid of challenges. Intentional inclusions like missing timestamps and speed anomalies mirror the complexities common in real-world telematics data. This design choice serves dual purposes:

\begin{enumerate}
  \item \textbf{Mimicking Real-World Scenarios:}  By simulating typical data challenges, our synthetic dataset presents a realistic backdrop. This allows us to genuinely assess and demonstrate ChatGPT's prowess in navigating these intricate issues.

  \item \textbf{Comprehensive Understanding:} The objective isn't merely to showcase how ChatGPT can aid in telematics data analysis but to understand the entire spectrum of the analytical process. By integrating stages like data cleaning and preprocessing, we paint a complete picture, emphasizing ChatGPT's versatility from the onset to the conclusion of the analysis.
\end{enumerate}

Given these underpinnings, the ensuing sections will elucidate how ChatGPT, as a robust tool, can be leveraged to address challenges endemic to telematics data.

\subsubsection{Removing or Filling Missing Values in Telematics Data}
Handling missing values is a fundamental step in the preprocessing of any dataset. For telematics data, the presence of missing values could stem from various reasons, including signal disruptions, sensor malfunctions, or transmission errors. Irrespective of the reason, missing values can significantly distort analyses, leading to biased, inconsistent, or even inaccurate outcomes. This is particularly critical in telematics, where data accuracy can directly impact decisions related to vehicle safety, maintenance, and operational efficiency.

There are various strategies to address missing values, depending on the nature and cause of the missingness, as well as the overall goal of the analysis. Some common approaches include:
\begin{enumerate}
  \item \textbf{Deletion:}  Removing rows with missing values. This is often the simplest approach but can result in significant data loss if missing values are widespread.
  \item \textbf{Imputation:} Replacing missing values with estimated values. This could be done using statistical methods (like mean, median, or mode imputation) or more sophisticated techniques such as interpolation, regression, or machine learning.
 \item \textbf{Forward or Backward Filling:} In time-series data like telematics, missing values can be replaced by preceding or succeeding values.
 \item \textbf{Flagging:} Creating a new variable to indicate when a value is missing, allowing for separate analysis of missing and non-missing data.
\end{enumerate}

Leveraging ChatGPT can simplify this process, providing quick and intuitive solutions tailored to the unique challenges of telematics data. Here, we will demonstrate how ChatGPT can be used to detect and handle missing values in our synthetic telematics dataset.

\begin{figure}[h]
\begin{center}
\includegraphics[width=145mm]{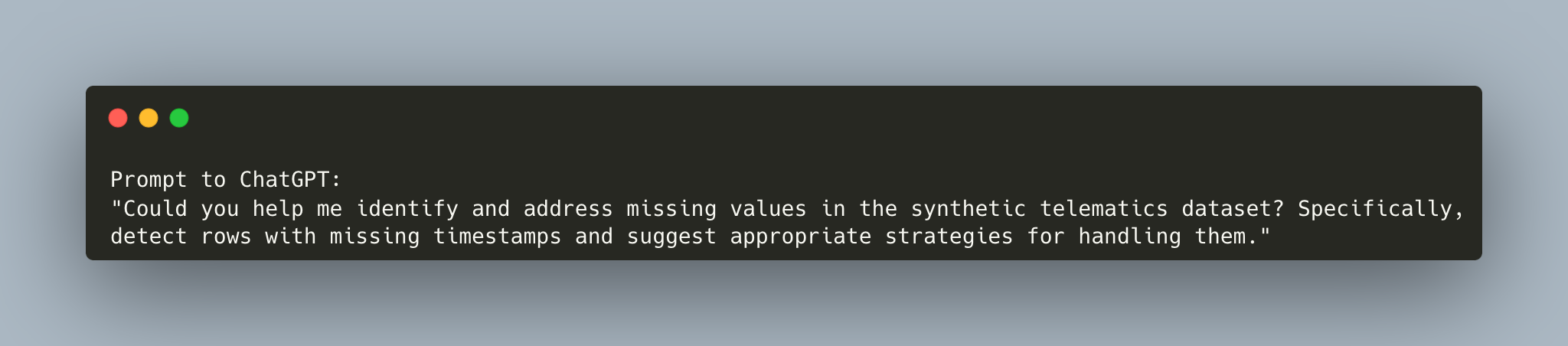}
\caption{Prompt to identify missing values.}
\end{center}
\end{figure}

\begin{figure}[h]
\begin{center}
\includegraphics[width=140mm]{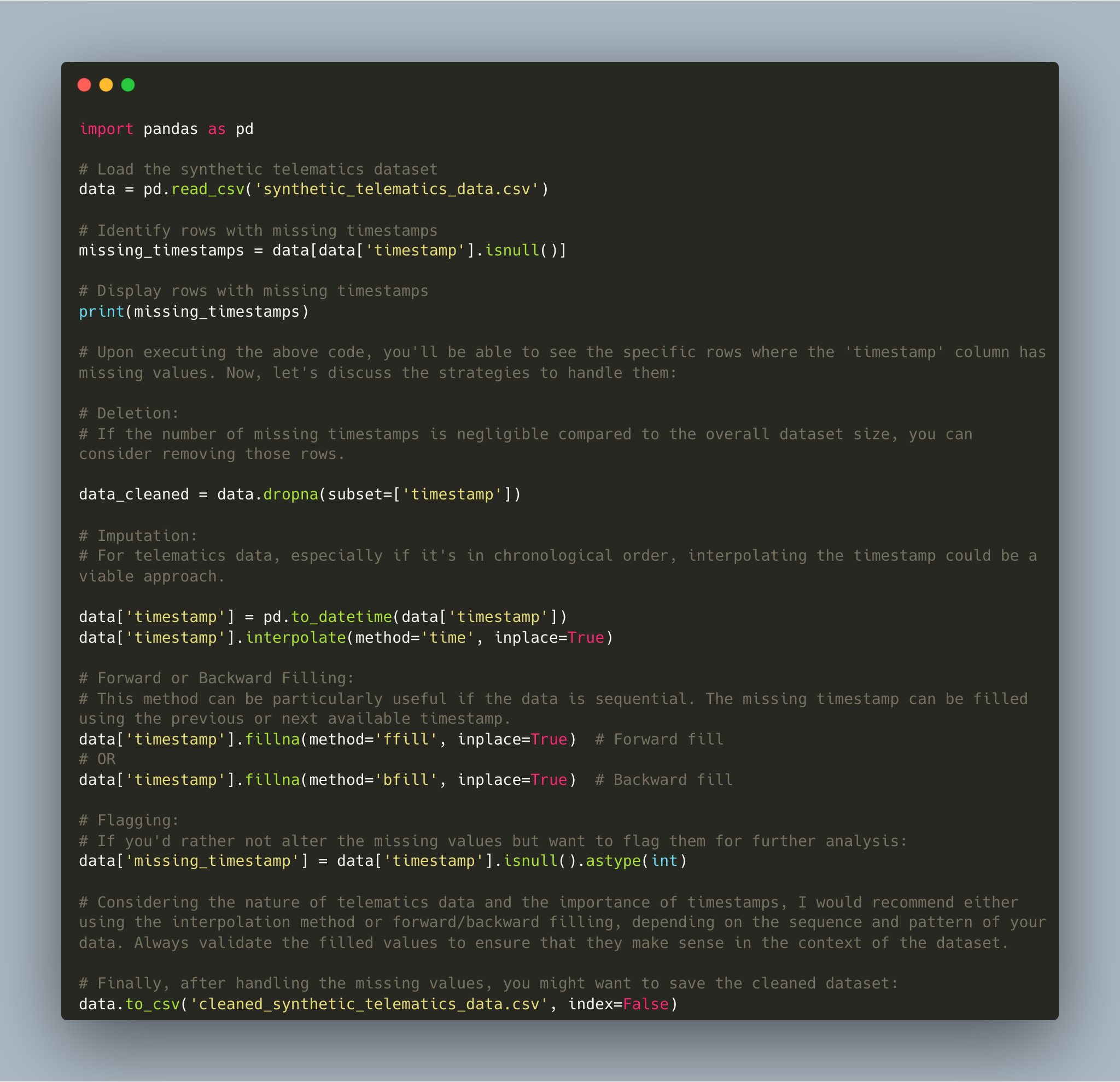}
\caption{ChatGPT output from missing values prompt.}
\end{center}
\end{figure}

In this subsection, we explored various strategies to manage missing timestamps in our synthetic telematics dataset. Given the dataset's structure and the intention behind introducing these missing values for educational purposes, we opted to delete the records with absent timestamps. By removing these entries, we maintain the integrity and continuity of the dataset, ensuring that subsequent analyses remain relevant and accurate.

While managing missing values is a common preprocessing task, it's just one aspect of data cleaning. Another fundamental step in preparing telematics data for analysis involves ensuring that each column or feature is of the appropriate data type. Incorrect data types can lead to erroneous computations and might hinder specific data manipulations.

\subsubsection{Converting Data Types with ChatGPT's Assistance}
Data type conversions are a common necessity in data preprocessing, especially when handling datasets from diverse sources or ones that have been merged from different origins. Proper data type management ensures accuracy in computations, eases data manipulation, and can even enhance memory optimization.

\textbf{Understanding the Current Data Types:}

Before attempting any type conversions, it's pivotal to understand the existing data types within the dataset. Typically, a quick overview with Python's Pandas library can provide this information:

\begin{figure}[h]
\begin{center}
\includegraphics[width=150mm]{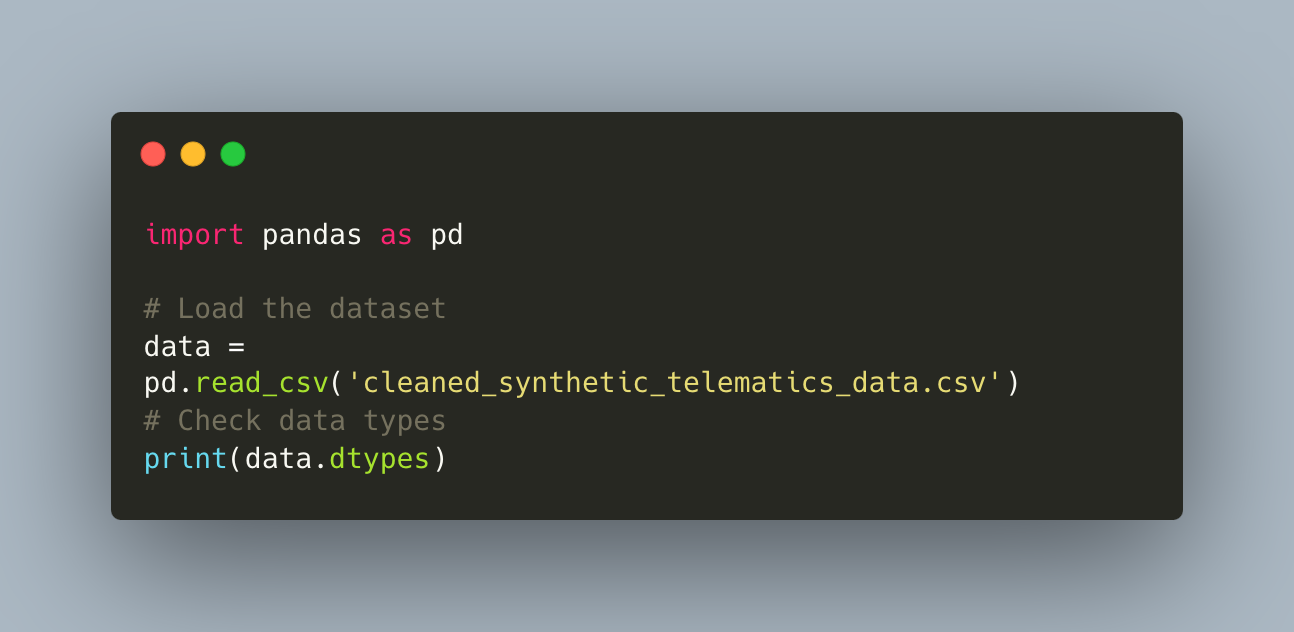}
\caption{ChatGPT code output to check data types.}
\end{center}
\end{figure}

\begin{table}[htbp]
\centering
\begin{tabular}{ll}
\textbf{Column} & \textbf{Data Type} \\
\hline
vehicle\_id & object \\
timestamp & object \\
start\_latitude & float64 \\
start\_longitude & float64 \\
end\_latitude & float64 \\
end\_longitude & float64 \\
average\_speed & int64 \\
\end{tabular}
\caption{Python output from running the above code.}
\label{tab:data_types}
\end{table}

\textbf{Identifying Required Conversions:}

Assuming a hypothetical output from the above script, let's say the 'timestamp' column is currently of object type (i.e., a string). It's more appropriate for this column to be in a datetime format for subsequent time series analysis or temporal computations.

\textbf{Converting Data Types:}

Below is the prompt and the resulting output it generated to demonstrate the conversion of the datetime data type. In the provided code snippet, ChatGPT used the pandas library to convert the 'timestamp' column in the DataFrame to datetime format. This conversion ensures that the values in the 'timestamp' column are treated as proper datetime objects, allowing for more efficient time-based operations and analysis. After executing the code, ChatGPT included an additional step to check the data types of the columns in the DataFrame. By printing the data types, we can verify the success of the conversion.

\begin{figure}[h]
\begin{center}
\includegraphics[width=150mm]{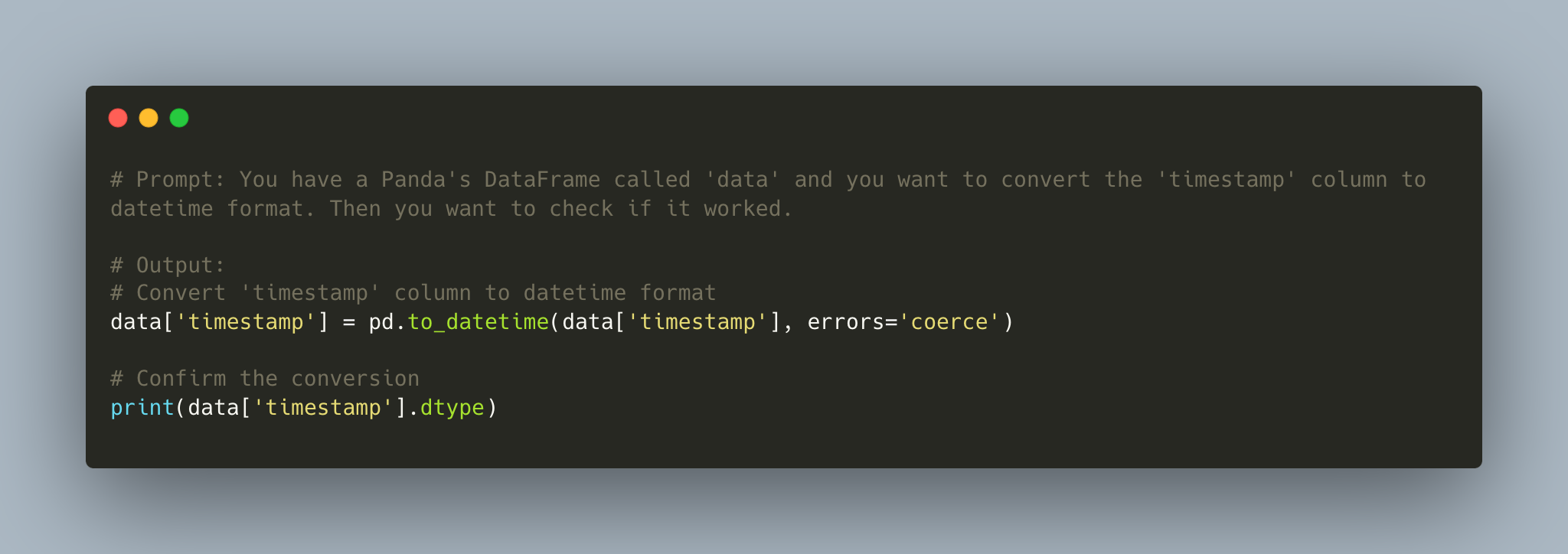}
\caption{Prompt to change the data type and ChatGPT code output.}
\end{center}
\end{figure}

Data type conversions might seem trivial, but they play a pivotal role in ensuring seamless data analysis. By ensuring each column is of its appropriate type, we can avoid unexpected errors, inaccuracies in computations, and facilitate further data manipulations.

With data types now appropriately managed, we can progress to another critical aspect of data preprocessing: outlier detection. Outliers can significantly distort statistical measures and analytics insights. Hence, the upcoming section will discuss leveraging ChatGPT in identifying and handling these anomalies within telematics data.

\subsubsection{Using ChatGPT to Help with Outlier Detection}
Outliers can pose significant challenges in the realm of data analysis. Especially within telematics, outliers can be indicative of transient glitches, errors in recording, or genuinely extreme events. By leveraging ChatGPT, researchers can enhance their methodologies in identifying and addressing such discrepancies.

Visual identification is a crucial preliminary step in the anomaly detection process. Particularly for large datasets where manual scrutiny is impractical, the ability to graphically represent data facilitates a quick understanding of its distribution and possible anomalies. In the domain of telematics, with vast streams of data from numerous vehicles, this visualization becomes even more pertinent.

\begin{figure}[h]
\begin{center}
\includegraphics[width=150mm]{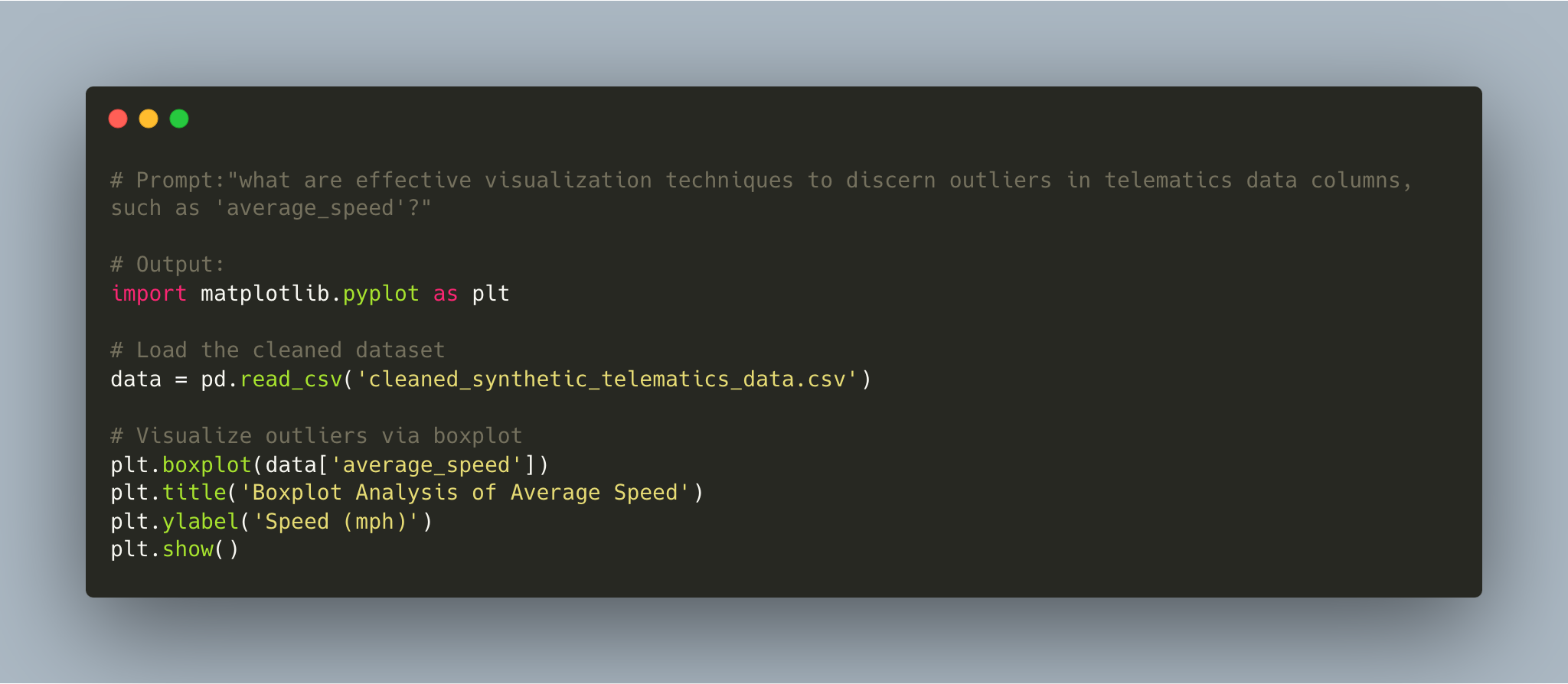}
\caption{Prompt to visualize outliers and ChatGPT code output.}
\end{center}
\end{figure}

Following the execution of this code, the output provides a boxplot visualization. Within this plot, the central box represents the interquartile range (IQR), highlighting the middle 50\% of the data, while the whiskers extend to encompass the majority of data points, barring outliers. Any points lying outside of these whiskers are typically treated as anomalies. This visual representation provides an immediate and clear understanding of the 'average\_speed' distribution and its outliers, which can then be further investigated.

% \begin{figure}[h]
% \begin{center}
% \includegraphics[width=120mm]{outlier_plot.png}
% \caption{Plot showing outliers generate from ChatGPT code output.}
% \end{center}
% \end{figure}

% Having visually identified these potential anomalies, please note the distinct circle around the 1000 mph mark in the box plot above, suggesting a clear deviation from the norm. To further corroborate this visual observation, it's instrumental to validate our findings with quantitative methods. One commonly employed statistical technique to detect outliers, especially in the context of telematics data given its often Gaussian distribution, is the Z-score. The ensuing section delves into how ChatGPT can guide users through the process of using the Z-score to affirmatively identify and manage these outliers.

\begin{wrapfigure}{l}{100mm} % 'l' indicates the image will be on the left
\centering
\includegraphics[width=100mm]{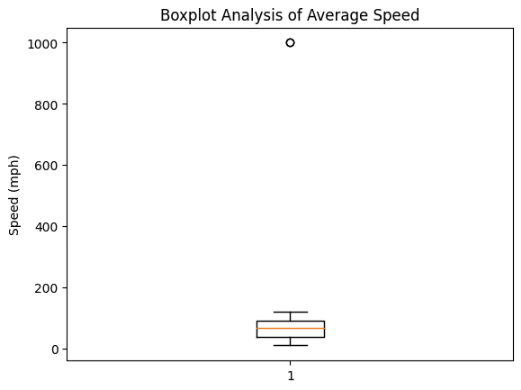}
\caption{Plot showing outliers generated from ChatGPT code output.}
\end{wrapfigure}

Having visually identified these potential anomalies, please note the distinct circle around the 1000 mph mark in the box plot to the left, suggesting a clear deviation from the norm. To further corroborate this visual observation, it's instrumental to validate our findings with quantitative methods. One commonly employed statistical technique to detect outliers, especially in the context of telematics data given its often Gaussian distribution, is the Z-score. The ensuing section delves into how ChatGPT can guide users through the process of using the Z-score to affirmatively identify and manage these outliers.

\textbf{Statistical Outlier Detection with ChatGPT's Assistance:}
Statistical approaches to anomaly detection are pivotal for several reasons. First, they offer a methodological foundation for identifying outliers, ensuring consistency and repeatability. Second, they can help confirm visual suspicions from plots and provide quantifiable measures for those outliers. Third, in scenarios where visualizations might be crowded or unclear, statistical methods can detect anomalies that might otherwise be missed.

Following this understanding, we proceeded to interface with ChatGPT, querying it about the application of Z-scores for anomaly detection in our dataset. Presented below is a stepwise depiction of our dialogue:

\begin{enumerate}
  \item \textbf{The Prompt:}  We initiated the conversation with ChatGPT by asking, "ChatGPT, how do statistical methods like Z-scores aid in outlier identification within datasets? Can you provide a Python example for its application?"

  \item \textbf{ChatGPT's Response:} ChatGPT provided a concise explanation and a Python code snippet demonstrating how to apply the Z-score methodology to identify outliers.

  \item \textbf{Implementation Outcome:} Subsequently, executing the provided Python code on our dataset enabled us to identify and validate the statistical outliers, particularly corroborating the visually detected anomalies around the value '1000' in the 'average\_speed' column.
\end{enumerate}

\begin{figure}[h]
\begin{center}
\includegraphics[width=120mm]{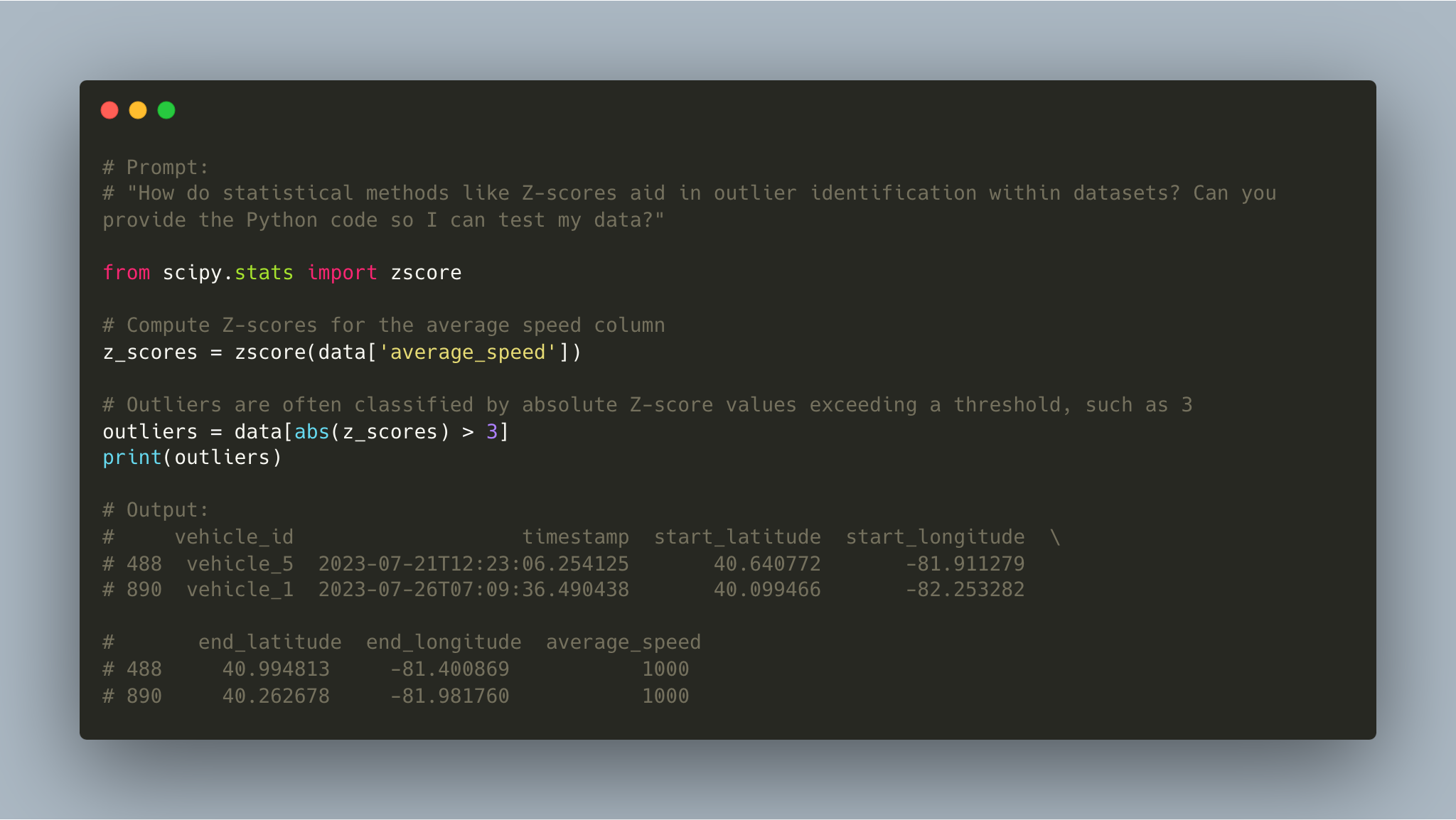}
\caption{Prompt, Code, and Output related to statistical outlier detection.}
\end{center}
\end{figure}

Having pinpointed the anomalies in our dataset, both visually and statistically, our next endeavor is to refine the dataset by addressing these outliers. Recognizing the skill of ChatGPT in data processing tasks, we intend to harness its capabilities to guide us through this filtering process. 

\textbf{Addressing Outliers Post-Detection:}

The subsequent step in this analytical journey involves determining the best course of action for these outliers. It's crucial to underscore that our dataset included these anomalies intentionally, as an educational exercise to walk readers through the process of detection and resolution. Thus, while identifying outliers is a pivotal phase, their appropriate resolution is the concluding task. Engaging ChatGPT with the prompt, "Having identified outliers in my telematics dataset, what are the academically accepted strategies for their resolution? And could you please provide the code to filter the outliers out." provides a broader perspective.

\begin{figure}[h]
\begin{center}
\includegraphics[width=130mm]{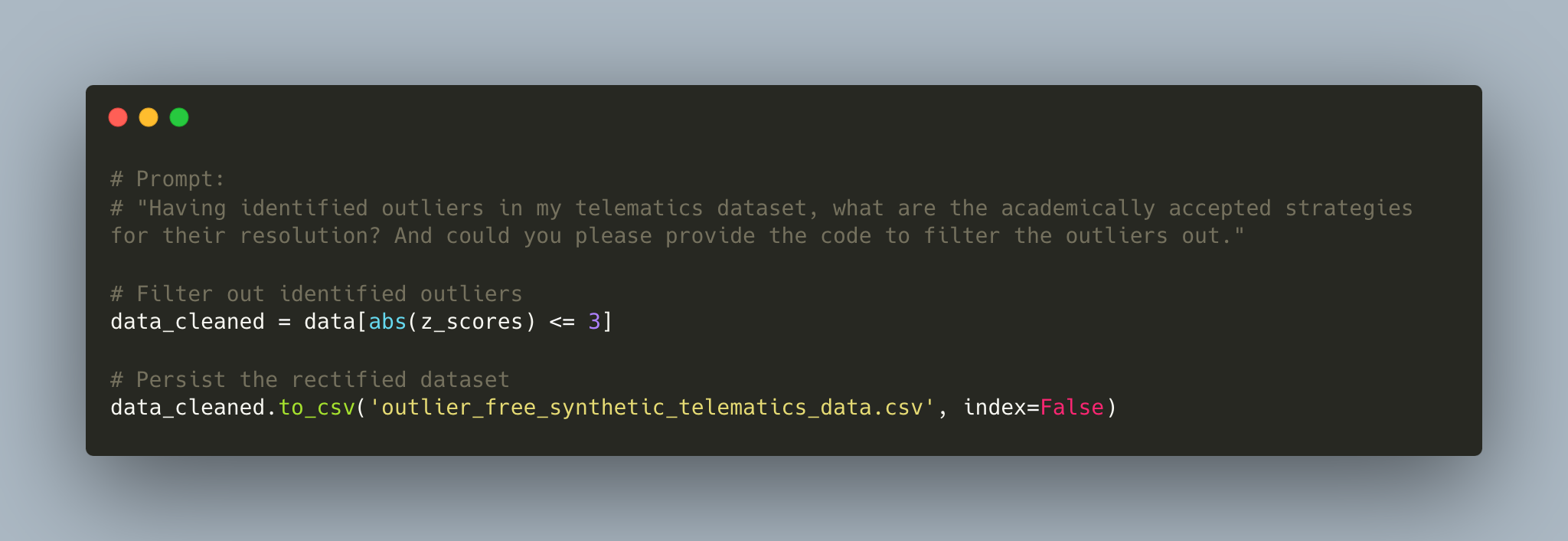}
\caption{Prompt and Code to filter out outliers.}
\end{center}
\end{figure}

In this section, we delved into the intricate challenges posed by telematics data, emphasizing the indispensability of preprocessing steps like data cleaning and outlier management. Through a constructive integration of ChatGPT into the analytical process, we showcased how the model can simplify tasks, offer guidance on best practices, and even provide actionable code snippets for efficient data processing. This rigorous cleanup ensures our dataset, though synthetic, closely mirrors the reliability and structure of real-world data, setting the stage for deeper insights and analyses.

As we transition to the subsequent section, it becomes imperative to acknowledge that data cleaning, while foundational, is just the beginning. The true essence of data analysis often lies in deriving newer insights by crafting meaningful features from the existing dataset. Feature engineering, therefore, stands as a cornerstone of effective data analysis, enabling the extraction of rich information and patterns otherwise latent. Join us in the next section as we explore how ChatGPT can be an invaluable ally in the realm of feature engineering, amplifying the depth and breadth of insights we can draw from our telematics data.

\subsection{Understanding Feature Engineering: A Key to Enhanced Data Analysis}
In the vast domain of data analysis, raw data, even after thorough preprocessing, may not always contain all the variables required to drive meaningful insights or optimal model performance. Feature engineering bridges this gap by generating new variables—features—from the existing data, offering a richer, more comprehensive perspective. It's the art and science of transforming raw data into features that better represent the underlying patterns to predictive models.

For telematics data, feature engineering can elevate the value of analyses significantly. Consider a scenario where we have start and end GPS coordinates of a vehicle journey. While these alone might not reveal much, derived features can offer deeper insights. For instance, calculating the direct distance traveled, or the average direction of a trip can unmask patterns about the efficiency of routes taken or habits of drivers. Such derived features greatly enhance our understanding and prediction capabilities.

In this section, with the help of ChatGPT, we will embark on deriving three key features for our dataset:

\begin{enumerate}
  \item \textbf{Day of the Week:}  Using the timestamp, we'll determine on which day of the week a particular trip occurred. This can reveal patterns such as higher travel frequencies on certain days or specific driving habits related to weekdays versus weekends.
  
  \item \textbf{Vehicle's Average Speed:} By analyzing the aggregate data, we'll derive a feature that captures the average speed for all trips taken by a vehicle. This can provide insights into general driving behavior or efficiency.

  \item \textbf{Distance Traveled:} From the start and end GPS coordinates, we will calculate the actual distance traveled during each trip, which can offer insights into the route efficiencies and the nature of trips (short errands versus long commutes, for instance).
\end{enumerate}
Leveraging ChatGPT in this phase can be immensely beneficial. The model can suggest methods to derive these features, assist in their computation, or even validate their relevance in a given context. Given its extensive training data and knowledge of best practices, ChatGPT can guide researchers in pinpointing features that might not be immediately obvious but can make a significant difference in analysis outcomes.

With this foundational understanding of feature engineering and our targeted objectives, let's delve into the practical aspects of generating these features with ChatGPT's guidance.

\subsubsection{Extracting the Day of the Week from Timestamps}
Within the field of telematics, features based on time frequently hold significant analytical value. Understanding not just when, but on which specific day an event or trip takes place, can lead to deeper insights. Such details can help organizations make sense of patterns like increased vehicular activity on specific days or shifts in driving habits between weekdays and weekends. Moreover, the day of the week could also correlate with certain driving behaviors, like increased speeds on weekends or more frequent stops during weekdays due to work-related commuting.

Given the sheer significance of this temporal dimension, extracting the day of the week from timestamps becomes indispensable. However, the raw timestamps, though rich in information, can sometimes be too granular for higher-level analyses. Therefore, a more abstract representation, like the day of the week, can be more suitable for certain analytical endeavors.

To achieve this transformation and bring this level of granularity into our synthetic telematics dataset, we can leverage ChatGPT's programming expertise.

\begin{enumerate}
  \item \textbf{The Prompt:}  We initiated the conversation with ChatGPT by asking, "ChatGPT,  I possess timestamps in the ISO 8601 format within my telematics dataset. How might I adeptly extract the specific day of the week from these timestamps using Python?"

  \item \textbf{ChatGPT's Response:} ChatGPT elucidated the importance of temporal features and furnished a Python code snippet detailing the process of extracting the day of the week from a timestamp.

  \item \textbf{Implementation Outcome:} Upon applying the shared Python code on our dataset, we successfully introduced a 'day\_of\_week' column. This newfound feature promises to be instrumental in subsequent analyses, revealing tendencies and behaviors distinct to each day.
\end{enumerate}

\begin{figure}[h]
\begin{center}
\includegraphics[width=130mm]{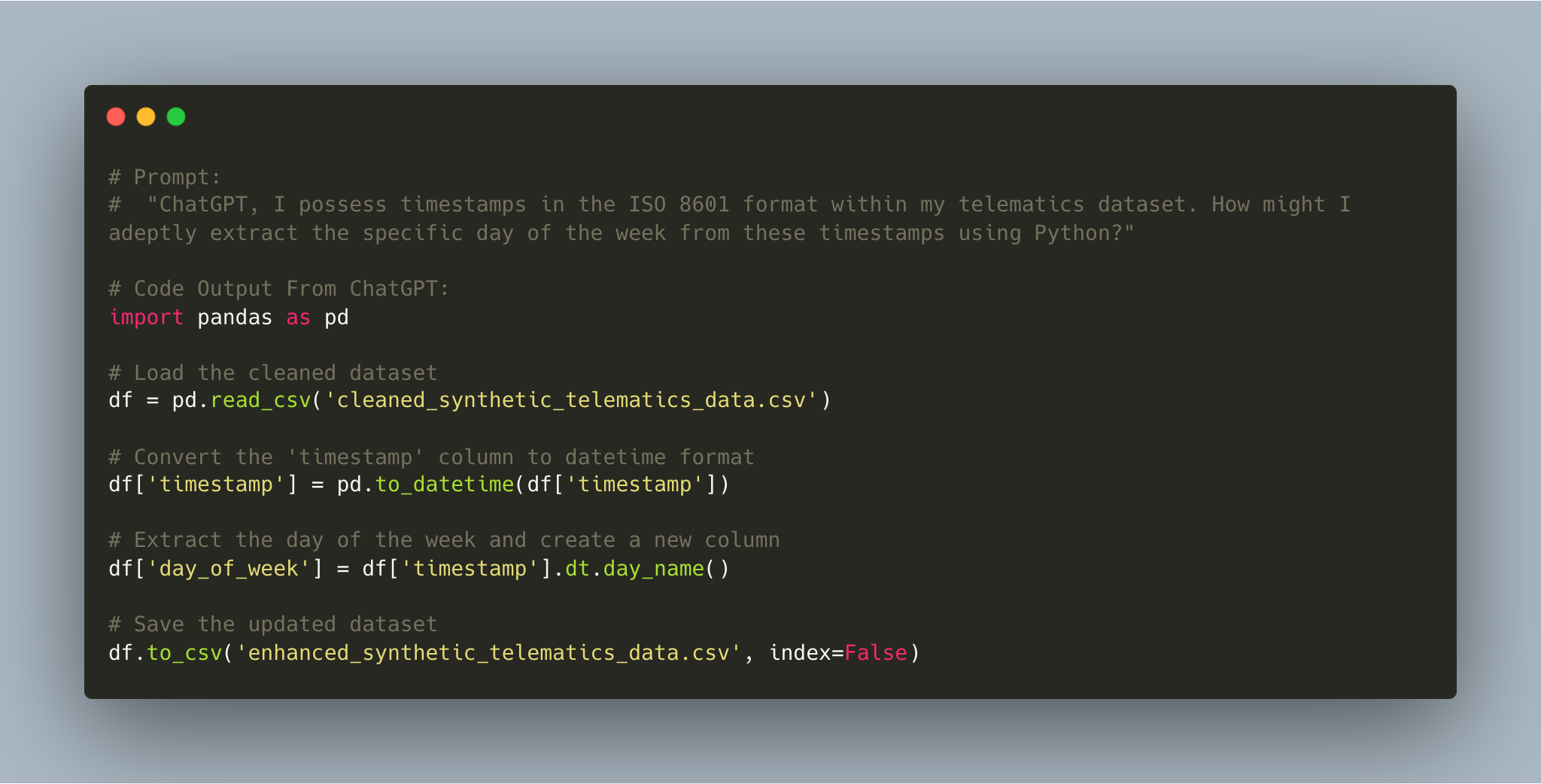}
\caption{Prompt and Code to add the day of the week column.}
\end{center}
\end{figure}

Having integrated the 'day\_of\_week' feature into our dataset, we can further enrich our analytical scope. The next step is to determine the average speed for each vehicle across all its trips, offering a broader perspective on vehicular behavior over time. Let's explore how ChatGPT can assist in crafting this aggregate metric.

\subsubsection{Computing the Average Speed for Each Vehicle Across All Trips}
Average speed is a fundamental metric in telematics, offering insights into a vehicle's typical operational patterns and potentially highlighting any systemic issues or inefficiencies. With a dataset that captures multiple trips for individual vehicles, the computation of this average provides a holistic view of vehicle behavior.

\begin{enumerate}
  \item \textbf{The Prompt:}  We initiated the conversation with ChatGPT by asking, "ChatGPT, I'm looking to calculate the average speed for each vehicle across all its trips in my dataset. Can you guide me with a Python example to achieve this?"

  \item \textbf{ChatGPT's Response:} To compute the average speed for each vehicle, you can group the dataset by the vehicle's identifier and then calculate the mean of the speed column for each group.

  \item \textbf{Implementation Outcome:} Upon executing the provided code, our dataset will now include an 'avg\_speed\_across\_trips' column for each vehicle. This aggregated metric will serve as a foundation for understanding broader driving patterns and ensuring consistent analytical comparisons between vehicles.
\end{enumerate}

\begin{figure}[h]
\begin{center}
\includegraphics[width=130mm]{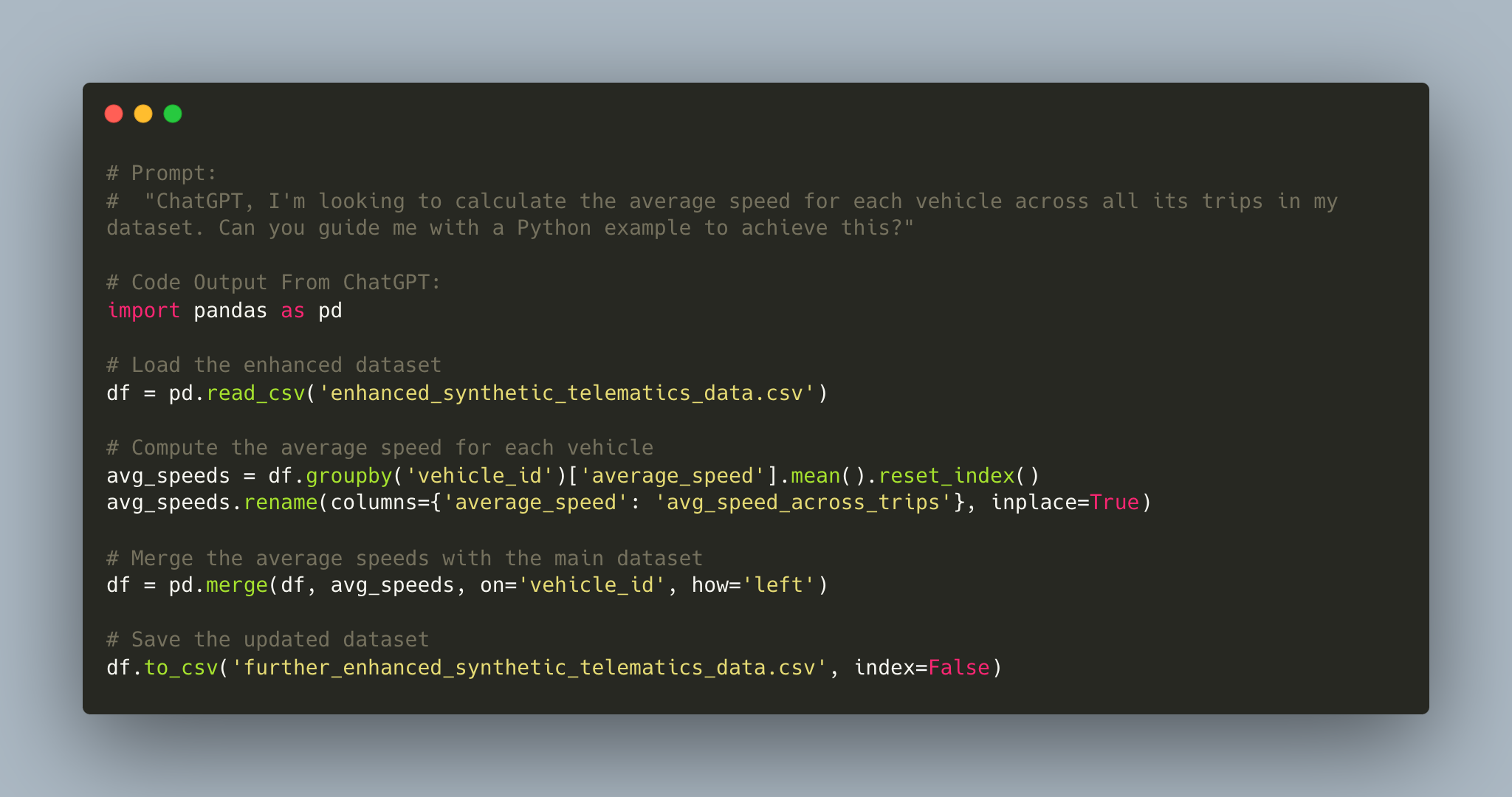}
\caption{Prompt and Code to create a column for total average speed.}
\end{center}
\end{figure}

Having deduced the average speed across trips for each vehicle, it becomes natural to further explore the depth of the data at our disposal. Next in our feature engineering journey is to compute another crucial metric: the total distance of each trip. Understanding trip distances can offer insights into usage patterns, the wear and tear on vehicles, or even the fuel efficiency based on varying trip lengths. Let's delve into this next task, aiming to calculate the total distance traveled during each trip.

\subsubsection{Calculating Total Distance Traveled for Each Trip}
Travel distance is an indispensable metric in telematics. By understanding the distance traveled in each trip, organizations can gain insights into various aspects such as fuel efficiency, wear and tear on the vehicle, and even predict the maintenance needs based on the miles logged. Moreover, patterns emerge, like certain routes being taken frequently or the typical distance covered by a vehicle on different days of the week.

To compute the total distance traveled during each trip, the dataset needs to provide either a series of GPS coordinates or a set of speed and time measurements. Given that our synthetic dataset includes speed values and timestamps, we can approximate the distance traveled over small time intervals and sum these to get the total trip distance.

\begin{enumerate}
  \item \textbf{The Prompt:}  We initiated the conversation with ChatGPT by asking, "ChatGPT, with start and end coordinates for vehicle trips in my dataset, how can I compute the total distance in miles traveled for each trip using Python?"

  \item \textbf{ChatGPT's Response:} ChatGPT recommended the haversine formula, a common method to ascertain the distance between two latitude-longitude points on the Earth's surface. The model provided a Python code snippet illustrating how this formula can be used to determine the distance between the 'start' and 'end' coordinates of each trip, outputting the result in miles.

  \item \textbf{Implementation Outcome:} Upon implementing the Python code given by ChatGPT on our dataset, we successfully added the total\_trip\_distance\_miles feature. This metric specifies the precise distance traveled in miles for every vehicle trip, offering a foundation for analyses such as evaluating routing efficiency and potential optimization avenues.
\end{enumerate}

\begin{figure}[h]
\begin{center}
\includegraphics[width=130mm]{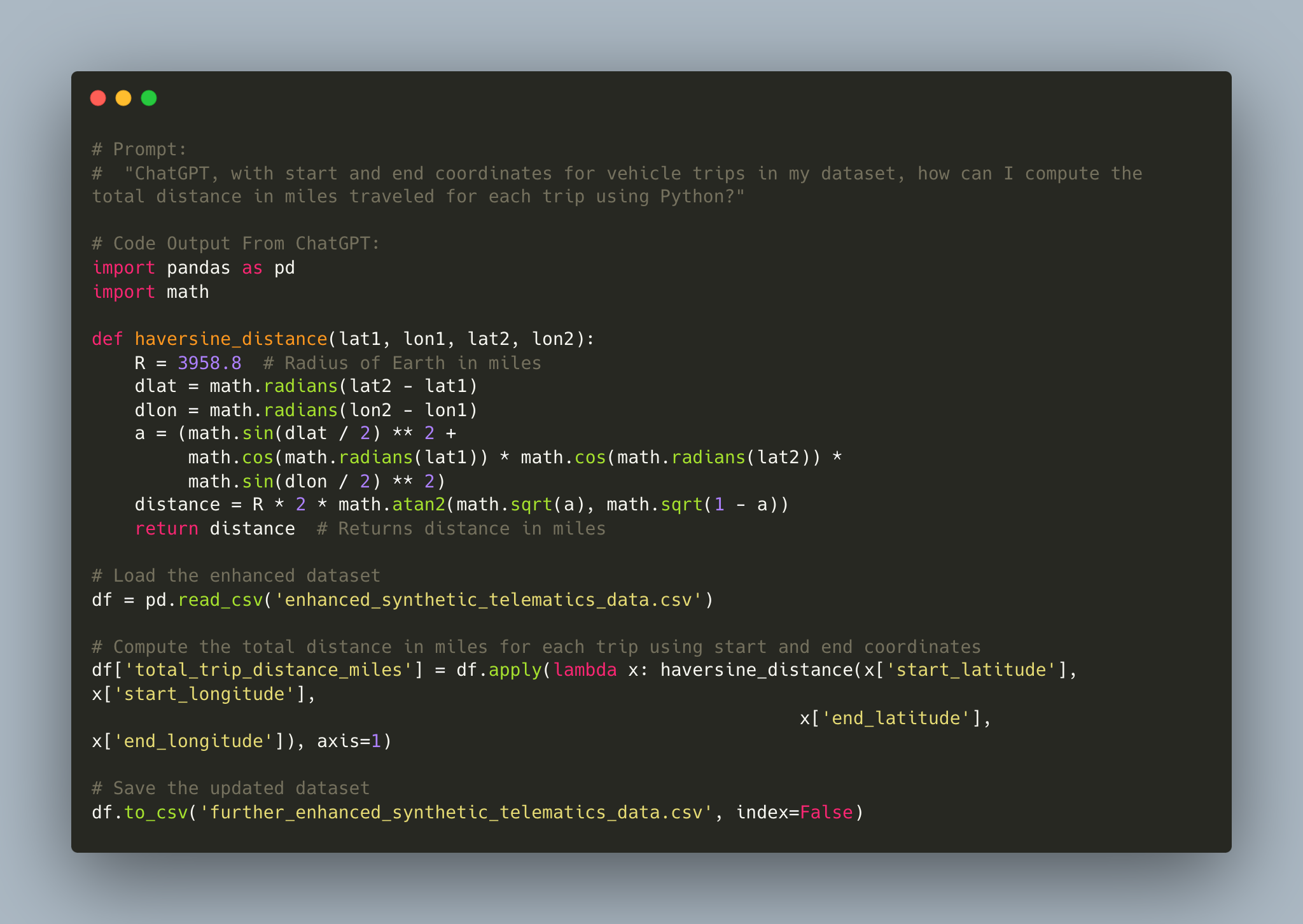}
\caption{Prompt and Code to create a column for total distance traveled during the trip.}
\end{center}
\end{figure}

With the addition of our new features, our synthetic telematics dataset is better prepared for analysis. Throughout the feature engineering process, ChatGPT was instrumental in providing both explanations and relevant code examples.

The inclusion of three key features: the day of the week, average speed, and total trip distance in miles, was primarily to demonstrate the process and show the reader how to leverage ChatGPT in real-world data analysis scenarios.

As we transition to the next section, our focus will shift to exploratory data analysis (EDA). EDA will help us understand the data's structure, outliers, patterns, and anomalies. With our enhanced dataset and the insights from ChatGPT, we're well-equipped for this analytical phase.

\subsection{Guided EDA with ChatGPT: Delving into Telematics Data}
\subsubsection{Introduction}

Our journey through the intricate landscape of data analysis has taken us through pivotal stages. Initially, we embarked on the foundational steps of data cleaning and preprocessing, ensuring the quality and coherence of our telematics data. This was succeeded by the phase of feature engineering, where we enriched our dataset with attributes such as the 'Day of the Week', 'Vehicle's Average Speed', and the 'Distance Traveled'. These enhancements enriched our dataset, equipping it with a depth that paves the way for nuanced analyses.

Now, as we transition into the Exploratory Data Analysis (EDA) phase, we're about to delve deeper into the data's story. EDA, a critical step in data analysis, enables us to uncover patterns, relationships, and insights that often remain obscured without this rigorous exploration. Given the multi-dimensional nature of telematics data, paired with the features we've crafted, EDA stands to offer a reservoir of findings.

Our exploration will begin by tapping into ChatGPT to generate Summary Statistics, capturing the central tendencies and variations within our telematics data. We will then move into Data Grouping, focusing on segmenting the dataset based on diverse criteria to uncover niche patterns. As we progress, Pattern Recognition will take the spotlight, aiding us in pinpointing correlations and intricate patterns within the dataset. Ensuring the reliability of our analysis, we'll venture into Anomaly Detection, identifying potential outliers or aberrations. Building on our findings, hypothesis testing will serve as a rigorous checkpoint, helping us validate or challenge our underlying assumptions. Lastly, our journey in this section will culminate with the visual representation of our EDA findings, where we'll translate the abstract numbers and patterns into intuitive and informative visuals.

With this roadmap in mind, let's dive into the heart of EDA, commencing with the essence of any data analysis, summary statistics.

\subsubsection{Summary Statistics}
The backbone of any initial data analysis lies in its summary statistics. It's akin to introducing oneself at the beginning of a conversation — a brief overview of who you are and what you bring to the table. Similarly, summary statistics provide a snapshot of our dataset's central tendencies, dispersions, and overall distribution. They offer a preliminary understanding, setting the stage for deeper investigations.

\begin{enumerate}
  \item \textbf{The Prompt:}  We initiated the conversation with ChatGPT by asking, "ChatGPT, considering our telematics data, how can we compute summary statistics to get a grasp on its central tendencies, dispersions, and distributions?"

  \item \textbf{ChatGPT's Response:} ChatGPT emphasized the importance of extracting measures like the mean, median, mode, standard deviation, and range. It then offered a Python code snippet for generating summary statistics.

  \item \textbf{Implementation Outcome:}Upon executing the provided code on our telematics dataset, we were greeted with a comprehensive table, as seen in Figure 15. It detailed the count, mean, standard deviation, minimum, 25th percentile, median (50th percentile), 75th percentile, and maximum values for each feature column. This quick overview was instrumental in giving us a bird's-eye view of our dataset's characteristics and potential areas of interest.
\end{enumerate}

\begin{figure}[h]
\begin{center}
\includegraphics[width=130mm]{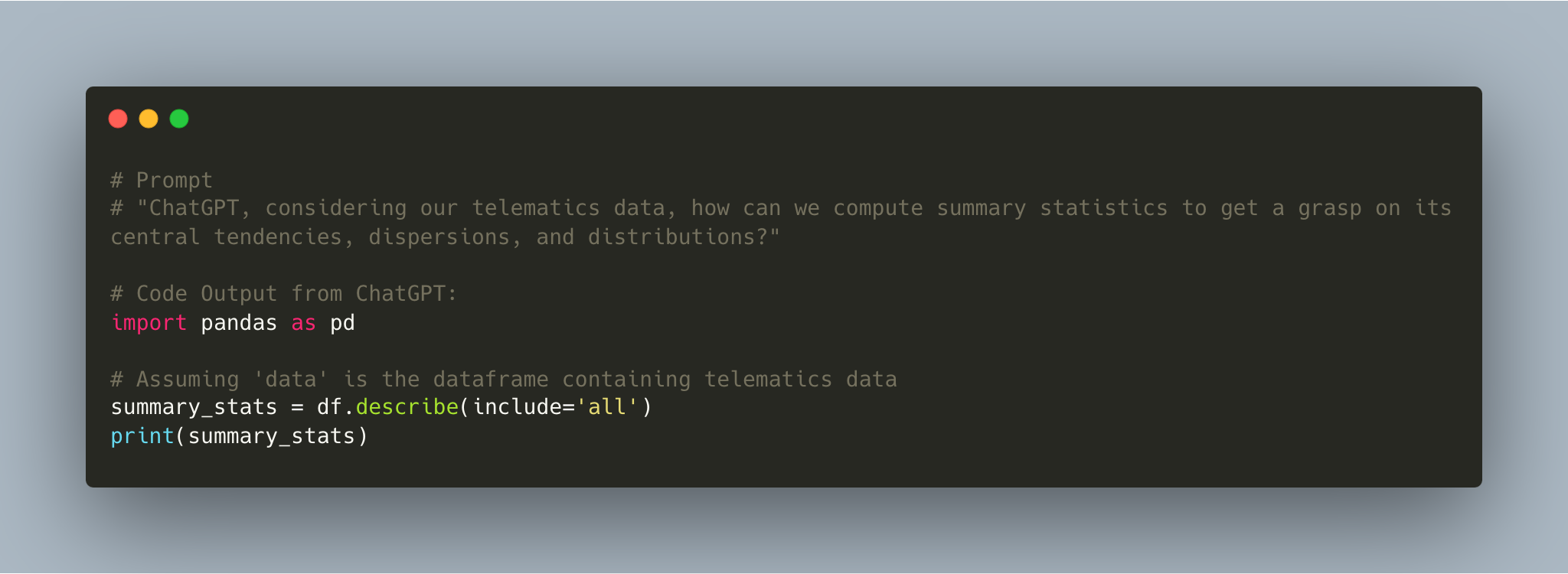}
\caption{Prompt and Code to summary statistics of data.}
\end{center}
\end{figure}

\begin{figure}[t]
\begin{center}
\includegraphics[width=130mm]{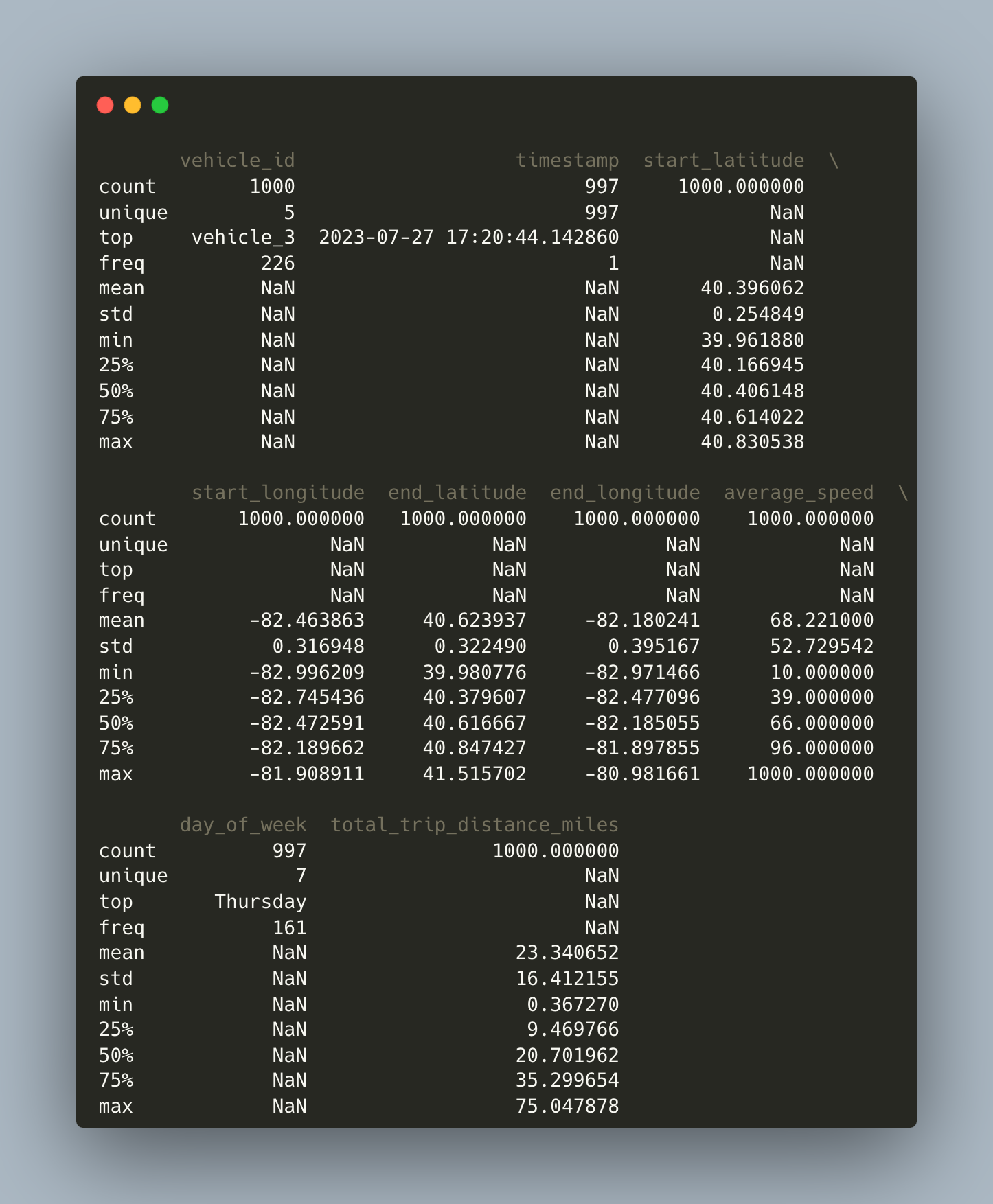}
\caption{Summary statistics output when you run the code from figure 14.}
\end{center}
\end{figure}

With a foundational understanding of our data in place through summary statistics, the picture starts to become clearer. Referring to the figures above and below, the figure above provides a concise display of our prompt to ChatGPT and its subsequent code response. Its simplicity and clarity are testament to how seamlessly artificial intelligence can integrate into data-driven workflows. The second figure below, a visual representation of the executed code, offers a tangible snapshot of our telematics data's overarching patterns. By studying this tableau of numbers, we gain insights into the general behavior of each feature, from their central tendencies to their variabilities.

Yet, as informative as summary statistics are, they merely scratch the surface. To extract the intricacies hidden within the data's layers, we must venture deeper, employing techniques that allow us to segment and dissect our data in varied dimensions. This segmentation can reveal more nuanced patterns and offer insights specific to different data subsets.

Energized with this preliminary knowledge, we gear up to dive into the next layer of our analytical journey: Data Grouping. Here, we aim to categorize our data based on distinct criteria, uncovering patterns that might remain concealed within a broader view.

\subsubsection{Data Grouping}
Data grouping or segmentation is a fundamental technique in data analysis that entails categorizing data into distinct subsets based on specific criteria. The goal is to dissect the broader dataset into manageable chunks, facilitating more targeted and granular analyses. Such segmentation can spotlight patterns, trends, or anomalies that might remain obscured when viewing the data as an undifferentiated whole.

In the context of telematics, a myriad of grouping possibilities exist: by geographical zones, vehicle types, driving behaviors, environmental conditions, and more. However, one of the most telling ways to segment this type of data is by time. The temporal aspect of telematics data reveals the dynamism and fluctuations inherent in vehicle and driver behaviors throughout the day.

For our specific exploration, we've chosen to focus on segmenting by time of day. The rhythm of daily life, marked by rush hours, work shifts, leisure activities, and resting periods, results in distinct driving patterns. By analyzing telematics data within these temporal confines, we seek to understand the ebb and flow of driving behaviors and how they correlate with specific times of the day.

\begin{enumerate}
  \item \textbf{The Prompt:}  We initiated the conversation with ChatGPT by asking, "ChatGPT, how can I segment telematics data by different times of the day to discern patterns related to driving behaviors and vehicle movements?"

  \item \textbf{ChatGPT's Response:} ChatGPT detailed a method for parsing our dataset's timestamps, categorizing them into specific time intervals such as morning, afternoon, evening, and night. It then provided a Python code snippet to facilitate this temporal segmentation.

  \item \textbf{Implementation Outcome:}  Upon executing the suggested code, our dataset was successfully segmented into distinct time-of-day categories. This allowed us to conduct more refined analyses, illuminating nuanced patterns associated with each time segment.
\end{enumerate}

\begin{figure}[h]
\begin{center}
\includegraphics[width=130mm]{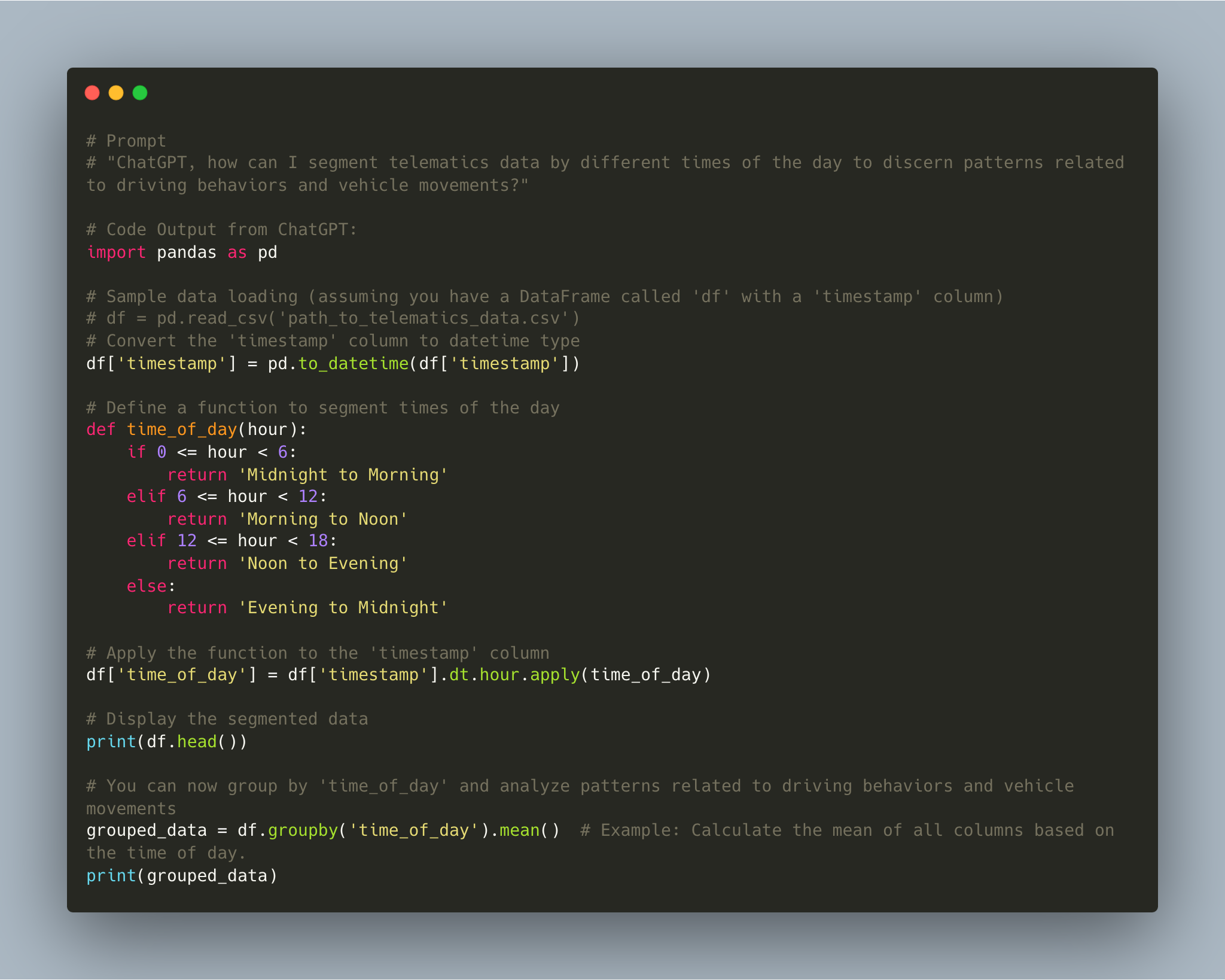}
\caption{Prompt and code to group data into different times of the day.}
\end{center}
\end{figure}
Segmentation in data analysis stands as a potent tool, enabling analysts to break down a vast pool of data into more manageable and interpretable chunks. By grouping our telematics data based on time intervals, we've unveiled nuanced insights into driving behaviors across different times of the day. Such granularity enhances our capacity to discern patterns, make predictions, and draw more specific, actionable conclusions.

As we've seen, time-based segmentation provides a prism through which we can view the multifaceted nature of telematics data. Whether it’s identifying rush hour traffic patterns or recognizing safer driving periods, segmentation equips us with a refined lens for analysis.

Having segmented our data and unlocked granular insights, our journey now advances to the realm of pattern recognition. Here, we'll be delving deeper into the data's intricacies, identifying correlations, and discerning relationships among variables. With the foundation laid by our data grouping exercise, we're poised to uncover patterns that would otherwise remain obscured in a sea of data. As we transition, let’s explore how ChatGPT can aid in demystifying these underlying patterns within our telematics dataset.

\subsubsection{Pattern Recognition}
Pattern recognition is an essential aspect of data analysis, especially when trying to discern intricate relationships and trends within the data. By identifying patterns, we can begin to understand behaviors, anticipate future occurrences, and even strategize interventions in various application areas.

Telematics data, in its rich and multifaceted nature, offers vast opportunities for pattern recognition. Factors such as geographical location, time, vehicle type, and driving behaviors intertwine to form intricate patterns that can shed light on numerous aspects, from efficient route planning to understanding peak travel times.

In this section, as a demonstrative example, we'll explore how the 'average speed' of vehicles might vary according to the 'time of day'. Such insights can be invaluable for transport and logistics planning or understanding rush hour dynamics.

It's essential to note, however, that the dataset we are working with here is somewhat simplistic. As datasets grow in complexity, the opportunities for pattern recognition increase exponentially. Our current endeavor is a 'toy' example, designed to illustrate the process and highlight how ChatGPT can assist. It's less about the depth of the patterns we might unearth and more about showcasing the methodology, offering a foundational approach for readers to then apply on more intricate datasets.

\begin{enumerate}
  \item \textbf{The Prompt:}  We initiated the conversation with ChatGPT by asking, "ChatGPT, using telematics data, how can I explore potential patterns between the average speed of vehicles and the time of day they travel?"

  \item \textbf{ChatGPT's Response:} ChatGPT emphasized that when dealing with relationships between a continuous variable like 'average speed' and a categorical variable such as 'time of day', visualization is paramount. It specifically recommended bar plots as they succinctly portray the average of a continuous variable across different categories. It further highlighted that while calculating correlations is beneficial for continuous-to-continuous relationships, in cases like ours, visual representation is more insightful.

  \item \textbf{Implementation Outcome:}  With guidance from ChatGPT, we generated a bar plot juxtaposing 'average speed' with distinct 'time of day' segments, as shown in figure 18. The visual portrayal unearthed some evident trends: for instance, a probable uptick in speeds during the 'Midnight to Morning' segment, potentially hinting at reduced traffic, and moderated speeds during busier times of the day. These visual insights, while not quantified with a correlation coefficient due to the categorical nature of the 'time of day', provided a profound understanding of the patterns inherent in our dataset.

\end{enumerate}

\begin{figure}[h]
\begin{center}
\includegraphics[width=130mm]{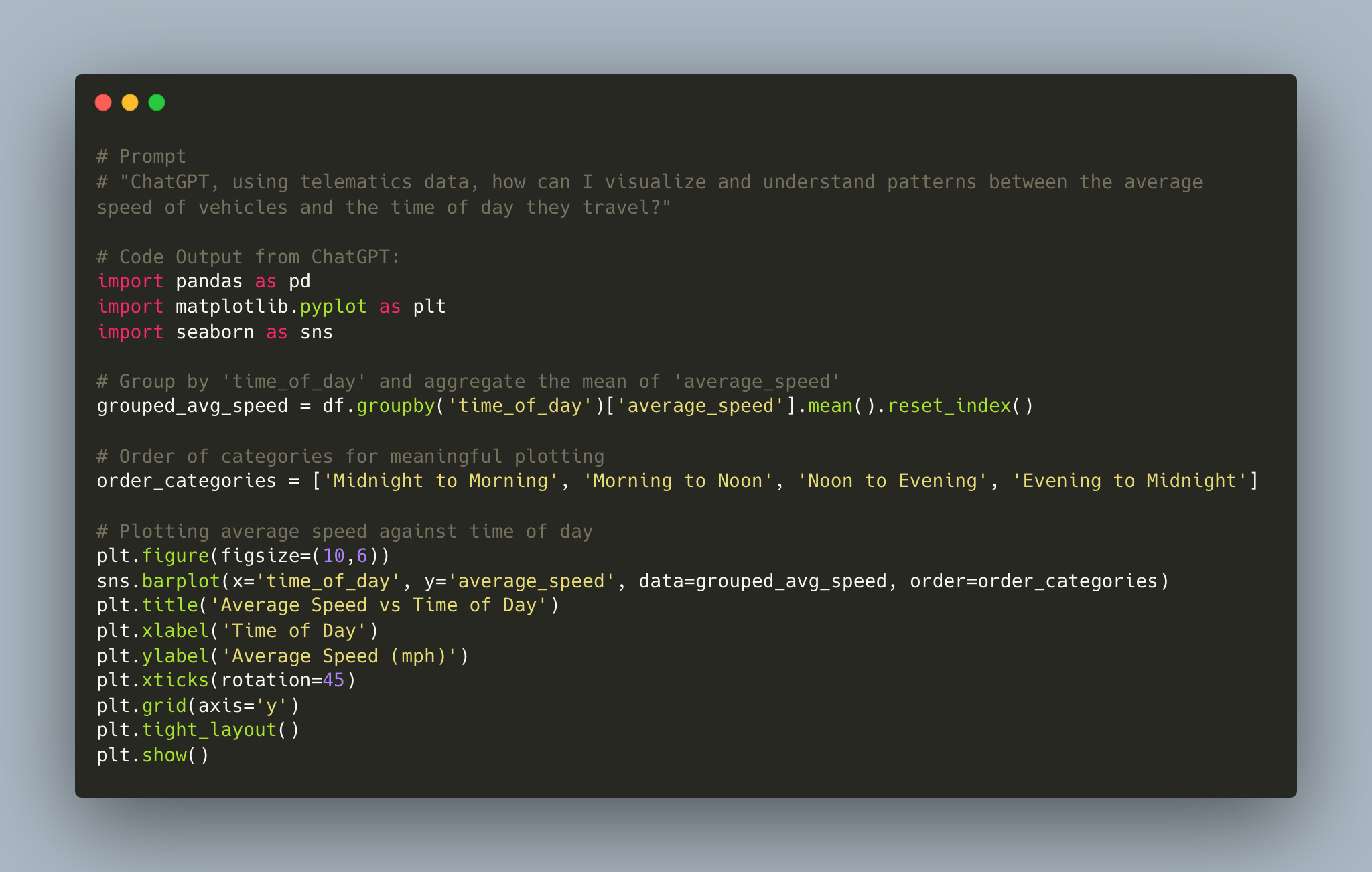}
\caption{Prompt and code to show the relationship between time of day and average speed.}
\end{center}
\end{figure}

\begin{figure}[h]
\begin{center}
\includegraphics[width=130mm]{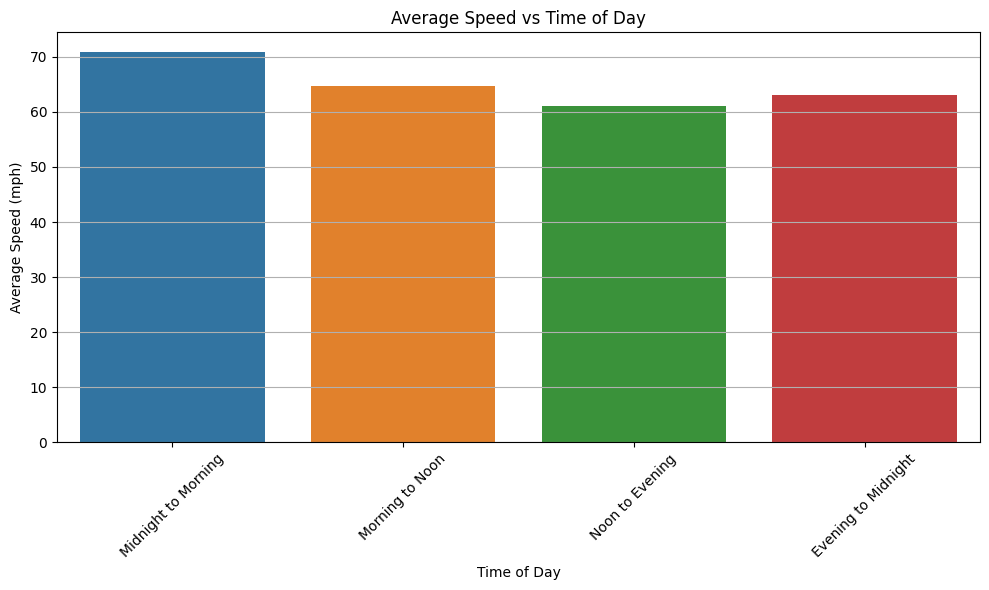}
\caption{Output from code in figure 17 that produces a bar plot showing the average speed for each time of day segment.}
\end{center}
\end{figure}

In the realm of telematics data, uncovering patterns is akin to decoding the language of vehicles and their drivers. Through our exploration of 'time of day' and 'average speed', we were able to discern driving habits and behaviors that varied with the day's progression. These insights, derived from a seemingly simple dataset, emphasize the power of structured analysis and visualization. They underline how even basic attributes, when correctly interrogated, can shed light on tangible trends that can inform everything from city planning to driver education programs. Again, to note, this data was synthetically generated. So, in a very real sense, there are no true insights. The goal of this paper is to show how to use ChatGPT to help generate the code needed to uncover insights when you are working with real data.

The ability to recognize patterns arms us with predictive capabilities, helping anticipate future trends or behaviors. Yet, just as important as spotting the norm is the capacity to identify the exceptions.

Patterns and trends give us a foundational understanding of the dataset. However, anomalies and outliers often reveal unexpected insights or possible data quality issues. As we advance to our next section, we delve into the realm of anomaly detection. Here, we will showcase how ChatGPT's help can be harnessed to spotlight these anomalies within our telematics dataset, highlighting potential areas of concern or interest.

\subsubsection{Anomaly Detection}
Anomaly detection is integral to data analysis, especially in fields where the outliers or deviations from the norm could have significant implications. In the context of telematics, anomalies could signify various scenarios - from data collection errors to unique driving behaviors or even potential vehicle malfunctions.

Recall, during our data cleaning process, we identified and addressed some anomalies, such as those trips with speeds clocking in at an implausible 1000 mph. Such outliers, if not detected, could severely skew our analyses. However, for the purposes of this section, we revert to our initial dataset to demonstrate the power and process of anomaly detection with the assistance of ChatGPT.

\begin{enumerate}
  \item \textbf{The Prompt:}  We initiated the conversation with ChatGPT by asking, "ChatGPT, given our initial telematics dataset, which might contain outliers like speeds of 1000 mph, how can we identify and analyze these anomalies?"

  \item \textbf{ChatGPT's Response:} ChatGPT outlined a multi-step approach to detect outliers:
\begin{enumerate}
  \item \textbf Start with visualizations like boxplots or scatter plots to visually identify any extreme values.

  \item \textbf Apply statistical measures such as the IQR (Interquartile Range) method or Z-scores to numerically determine anomalies.

  \item \textbf Once detected, further analyze these anomalies to determine their cause, and decide whether they should be removed, corrected, or kept for further study.

\end{enumerate}

  \item \textbf{Implementation Outcome:}  Leveraging ChatGPT's guidance, we began by visually representing the 'average speed' variable using a boxplot. This visualization technique immediately drew attention to the stark outliers present in the data. To further refine our understanding, we employed Z-scores, which facilitated the precise identification of anomalous data points based on their deviations from the mean speed.

\end{enumerate}

\begin{figure}[h]
\begin{center}
\includegraphics[width=130mm]{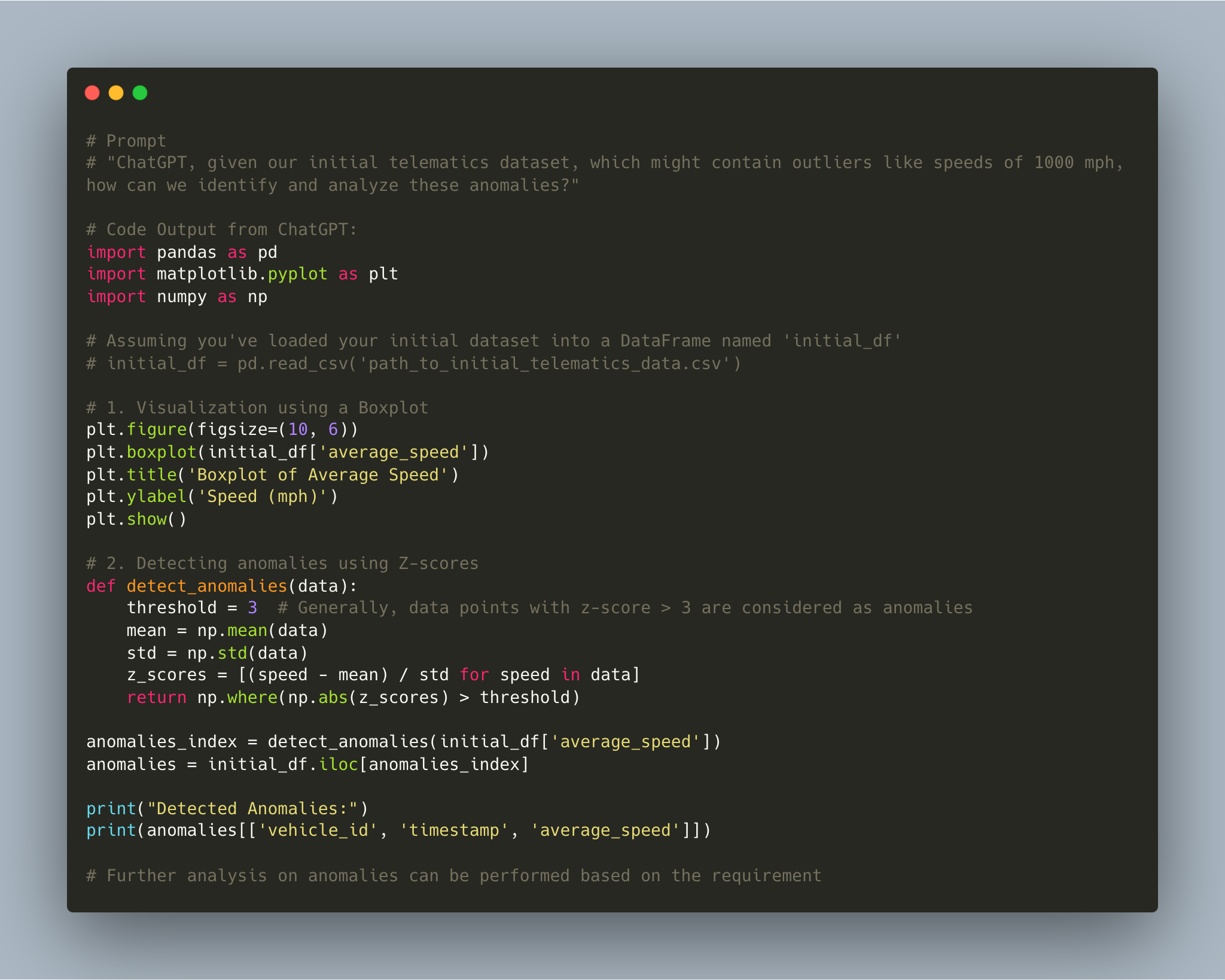}
\caption{Prompt and code to show anomolies.}
\end{center}
\end{figure}

In revisiting our dataset, we recognized these anomalies, which we had purposefully introduced as extreme outliers, to illustrate the importance and methodologies of anomaly detection. This exercise reiterated the necessity of a meticulous data cleaning regimen. 

Anomalies, while they deviate from the norm, are vital aspects of our data narrative. Whether they indicate data quality issues, rare events, or significant occurrences, recognizing and understanding these outliers is crucial. Through visual and statistical methods, we not only identified but also delved deeper into the underlying causes of these anomalies, showcasing the depth and rigor required in comprehensive data analysis.

Having identified patterns and outliers within our dataset, we now move to a stage that builds on these foundational insights. The upcoming section will take us into the realm of hypothesis testing, where we leverage ChatGPT to draw inferences and test assumptions, further enriching our understanding of the telematics data.

\subsubsection{Hypothesis Testing}
Hypothesis testing is a foundational technique in statistical analysis. It enables us to make informed inferences about our data by comparing different groups or evaluating potential relationships between variables. Particularly in telematics, where vast volumes of data can sometimes obscure meaningful differences, hypothesis testing serves as a robust tool to discern whether observed patterns or relationships are statistically significant or merely due to chance.

Given the diverse nature of telematics data, myriad hypotheses could be posited. For the purposes of this analysis, we'll focus on a simple yet illustrative example. Let's test if the average speed of vehicles during "Morning to Noon" significantly differs from that during "Evening to Midnight".

\begin{enumerate}
  \item \textbf{The Prompt:}  We initiated the conversation with ChatGPT by asking, "ChatGPT, could you guide us through conducting a hypothesis test to compare the average speeds of vehicles during 'Morning to Noon' and 'Evening to Midnight' in our telematics dataset?"

  \item \textbf{ChatGPT's Response:} ChatGPT outlined the steps for performing a two-sample t-test, given that we're comparing the means of two independent groups. It emphasized checking the assumptions of normality and homogeneity of variances before proceeding. Furthermore, it suggested calculating the p-value to determine the statistical significance of the observed differences.

  \item \textbf{Implementation Outcome:}  Following ChatGPT's guidance, we performed a t-test on our dataset after ensuring normality and homogeneity of variances. The resulting p-value from the test confirmed that there was no statistically significant difference in average speeds between the two time segments.
\end{enumerate}

\begin{figure}[h]
\begin{center}
\includegraphics[width=130mm]{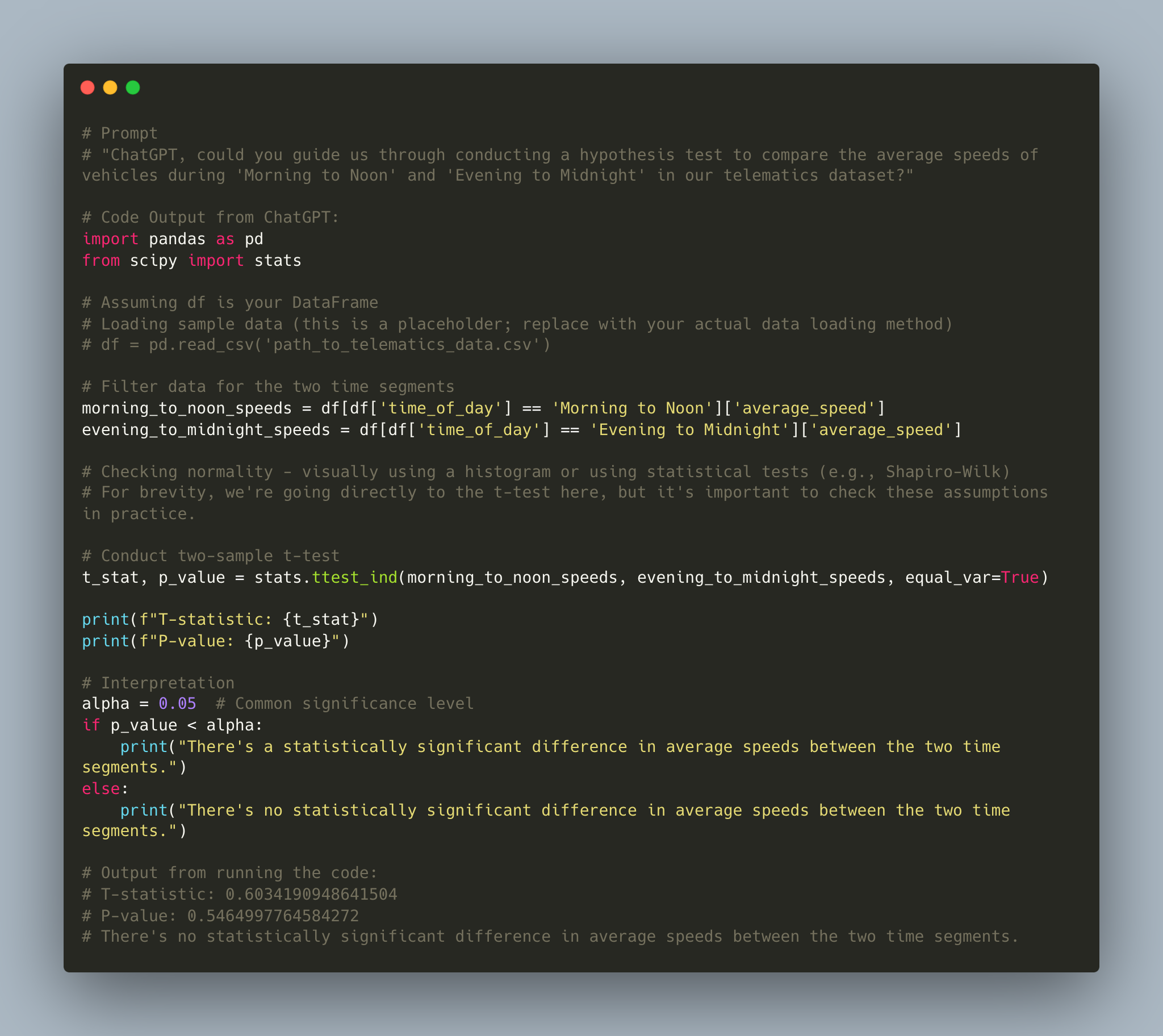}
\caption{Prompt, code, and output for the hypothesis test.}
\end{center}
\end{figure}

Having understood the significance of observed patterns using hypothesis testing, visualization remains a key aspect of effectively communicating these insights. As we move forward, we'll explore how ChatGPT assists in visually representing the EDA findings, making them more interpretable and impactful for our audience.

\subsubsection{Visual Representations}
Visual representation has been an integral aspect throughout our research and analysis presented in this paper. As we've navigated the landscape of our telematics dataset, graphs, charts, and plots have provided us with intuitive and immediate insights. The power of visualization lies in its ability to make complex data relationships understandable, aiding not just in pattern recognition but also in effectively communicating those patterns to others.

As we conclude our exploration of telematics data with ChatGPT, it's worth emphasizing a crucial point: EDA is not strictly linear. It's iterative and cyclical. Just as you might loop back to data cleaning after spotting anomalies during visualization, you might also find that an unexpected pattern in your data pushes you to generate new features or segment your data in a different way. It's a dynamic dance of inquiry, driven by both intuition and the insights that emerge at each stage.

In this section, while we've already integrated visualization techniques throughout our analysis, we aim to showcase a few more advanced graphical techniques. This will further highlight the versatility of ChatGPT in assisting with diverse visualization needs.

\begin{enumerate}
  \item \textbf{The Prompt:}  We initiated the conversation with ChatGPT by asking, "ChatGPT, now that we've explored the telematics data extensively, could you suggest advanced visualization techniques that can best encapsulate the myriad of insights we've uncovered?"

  \item \textbf{ChatGPT's Response:} Visual representation is a crucial facet of any data exploration journey. For the diverse insights from your telematics data, I recommend the following advanced visualization techniques:
  
  \begin{enumerate}
  \item \textbf {Heatmaps:} Especially useful for showcasing correlations between multiple variables in a grid format. Given your analysis of patterns and anomalies, a heatmap might provide clarity on the intensity and relationship of those variables.

 \item \textbf {Violin Plots:} Combining elements of box plots and density plots, they provide a deeper understanding of the distribution of continuous data against different categories. For instance, comparing 'average speed' across different 'times of the day' can yield nuanced insights.

 \item \textbf {Pair Plots:} A matrix of scatter plots to visualize pairwise relationships between several variables simultaneously. Given the range of features in your dataset, pair plots can help identify intricate patterns quickly.

 \item \textbf {Facet Grids:} Enables the creation of a grid of plots based on values of one or more categorical variables. For instance, examining 'average speed' trends across different days of the week, segmented by 'time of day'.

 \item \textbf {Interactive Plots:} Using libraries like Plotly or Bokeh, you can generate dynamic visualizations. This is especially useful when dealing with a multifaceted dataset, as viewers can zoom, pan, or hover to get detailed information.

\end{enumerate}

  \item \textbf{Implementation Outcome:} Acting on ChatGPT's advice, we delved into crafting these visualiztions:
   \begin{enumerate}
  \item \textbf {Heatmaps:} helped us perceive correlations, particularly between 'average speed' and 'time of day', with colors intensifying during specific hours, hinting at patterns we previously identified.

 \item \textbf {Violin Plots:} illustrated the density and distribution of 'average speed' across different 'times of the day', revealing not just the central tendency but also the variability in speeds.

 \item \textbf {Pair Plots:} provided a snapshot of relationships across multiple variables, offering a holistic view and pointing to potential areas of further investigation.

 \item \textbf {Facet Grids:} discerned patterns for 'average speed' for each 'day of the week', segmented by 'time of day'. This granular visualization further confirmed patterns and anomalies in driving behavior.

 \item \textbf {Interactive Plots:} engage dynamically with the data, zooming into areas of interest, or hovering over data points for more details provided an immersive experience, making the insights accessible and compelling.

\end{enumerate}

\end{enumerate}

Following ChatGPT's suggestions, we embarked on a visualization journey. We began with heatmaps which illuminated the strong correlations in our dataset. Particularly, the association between 'average speed' and 'time of day' stood out. As the colors intensified during specific hours, it reemphasized patterns we had identified earlier.

Our exploration led us to violin plots. These plots beautifully showcased the density and distribution of 'average speed' throughout different 'times of the day'. They not only highlighted the central tendency of speeds but also shed light on the variability, painting a vivid picture of driving behaviors.

To understand relationships across multiple dimensions, pair plots came to our aid. These plots offered an integrated perspective, intertwining relationships across variables and spotlighting potential focal areas for a deeper dive.

We also adopted facet grids to get a more segmented view. These grids allowed us to observe 'average speed' patterns distinctly for each 'day of the week' and within the subsets of 'time of day'. This granularity reaffirmed our understanding of driving behavior, and the visual cues helped anchor our insights.

Concluding our visualization odyssey, we turned to interactive plots. The ability to delve into specific areas of the plot, zooming in and out, or gleaning more details by simply hovering over data points transformed our data exploration into an engaging experience, making the insights both accessible and striking.

Presented below in Figure 21 is the prompt we issued to the model, along with the corresponding code it generated. This code exemplifies the application of advanced visualization techniques, as recommended by ChatGPT, to unveil deeper insights from our telematics dataset.

\begin{figure}[h!]
\begin{center}
\includegraphics[width=130mm]{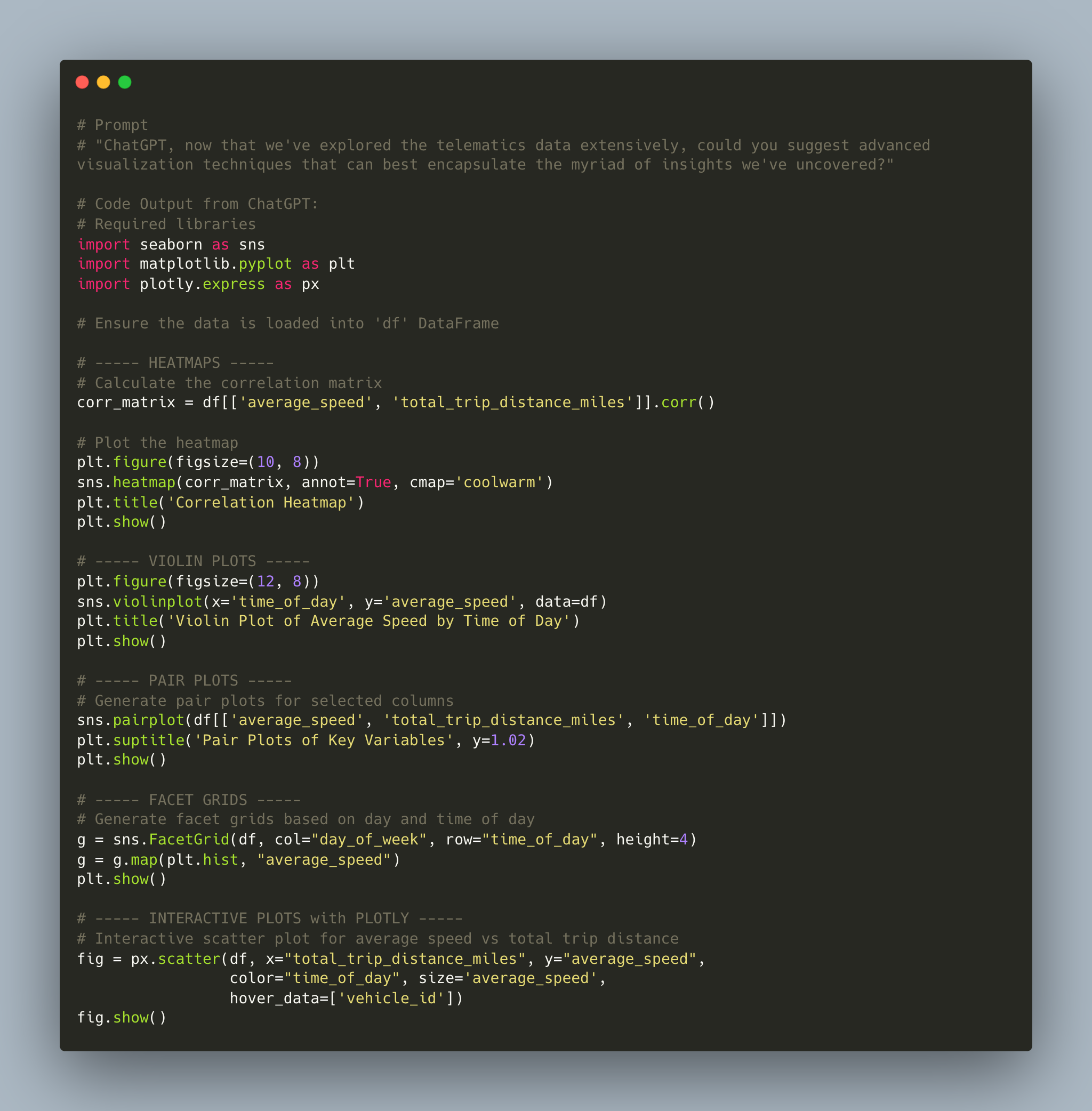}
\caption{Prompt and code to generate visualizations.}
\end{center}
\end{figure}

\subsubsection{EDA Conclusion}
The journey through EDA, with guidance from ChatGPT, has been enlightening. We embarked on this expedition by understanding the overarching structure of our dataset, ensuring we had a clear and holistic view. Segmentation further enhanced our comprehension, allowing for granulated, targeted insights. We then delved into pattern recognition, seeking the inherent rhythms and regularities within our data and contrasting them with the anomalies, those critical deviations that often hold the most intrigue.

The process highlighted the need for meticulous examination, signaling possible data inconsistencies and solidifying trust in our ensuing evaluations. Furthermore, the myriad of visual representations we employed resonated with the age-old sentiment: "A picture is worth a thousand words." This assortment of charts and diagrams elegantly illustrated intricate interconnections, rendering multifaceted data relationships both insightful and readily comprehensible.

In essence, EDA, as facilitated by ChatGPT, provided a roadmap. It laid the groundwork, ensuring that subsequent analyses and models built on a robust, well-understood foundation. It's a reminder that before diving deep into the intricacies of data science and analytics, one must first appreciate, understand, and navigate the vastness and complexity of the data landscape.

\section{Democratizing Data Science}
\subsection{Introduction}

In an ever-evolving data landscape, the term "democratization" signifies making data analysis more accessible to a broader audience, mitigating technical expertise barriers. We're on the brink of a transformative era where artificial intelligence meets natural language processing. This convergence, exemplified by tools like ChatGPT, is revolutionizing data science. By blending AI and NLP, ChatGPT is paving the way for a more inclusive and democratic approach to data analysis, opening doors for a wider audience.

\subsection{Lowering Barriers to Entry for Beginners through Natural Language Interface}
The traditional learning curve for data science can be intimidating, steeped in the complexities of programming languages, intricate statistical concepts, and specialized software tools. A common deterrent for many potential data enthusiasts is the often daunting syntax of programming languages like Python, R, or Julia.

However, with the rise of natural language interfaces, exemplified by tools like ChatGPT, we're witnessing a transformative shift. As this paper has shown, instead of wading through dense lines of code, users can engage with data using plain, conversational English. This intuitive nature of AI-powered interfaces, such as ChatGPT, equips beginners to concentrate on the essence of data analysis — framing the right questions and interpreting insights, without the distractions of technical hurdles.

Imagine a novice wanting to discern patterns in a dataset. Rather than grappling with the intricacies of a Python script, they could simply ask, "ChatGPT, can you help me identify trends in this data?" This direct, jargon-free approach is not only more accessible but also positions the individual closer to the core of data exploration.

By minimizing these initial barriers, tools like ChatGPT pave the way, guiding beginners from a casual curiosity in data to a more profound and informed engagement with data science.

\subsection{Shifting Focus: From Complex Programming Syntax to Problem-Solving and Analysis}

Historically, data science has been as much about mastering programming as it has been about understanding the nuances of the data itself. The process of deriving meaningful insights often became overshadowed by the challenge of wrangling with intricate code, troubleshooting errors, and optimizing algorithms. This syntactical challenge would often be a primary obstacle, diverting attention away from the actual analytical task at hand.

As demonstrated in the previous sections of this paper, ChatGPT offers a glimpse into a future where the emphasis is refocused. The integration of natural language interfaces into data science tools shifts the paradigm. Instead of spending a significant portion of time understanding and crafting the perfect line of code, individuals can now pose questions, hypothesize, and engage in direct analytical thinking.

This shift is profoundly democratizing. It means that domain experts without formal programming training, whether they are economists, biologists, or historians, can now directly interrogate datasets pertinent to their fields. It fosters an environment where the emphasis is on the 'science' in data science --- the formulation of hypotheses, the exploration of data, and the interpretation of results.

By streamlining the analytical process and removing syntactical distractions, we're evolving towards a landscape where the core of data science --- the quest for understanding and insight --- becomes universally accessible.

\subsection{The Impact of Rapid Feedback from ChatGPT on Speed and Efficiency of Data Analysis}

A distinguishing factor in the use of ChatGPT in data science, as highlighted throughout our exploration, is the rapid feedback loop it provides. Traditional data analysis often involves writing code, executing it, interpreting the output, and then iterating on the code again. This cycle, while effective, can be time-consuming, especially when faced with errors or unexpected results.

With ChatGPT, data queries and explorations are executed in near-real-time. The instantaneity of feedback accelerates the analytical process. Immediate responses not only save time but also keep the analyst's train of thought undisturbed, enabling deeper and more consistent immersion into the analysis.

For beginners, this rapid feedback serves as an efficient learning tool. Mistakes or misconceptions are addressed promptly, allowing for quick course corrections and enhanced understanding. For seasoned data scientists, it streamlines exploratory phases, helping them rapidly sift through hypotheses or test assumptions.

In essence, the immediate feedback from ChatGPT acts as a catalyst. It fosters a dynamic environment where insights are gleaned faster, learning curves are shortened, and the overall efficiency of the analytical process is elevated.

\subsection{Democratizing Data Science for Non-tech Domain Experts}
The potential of data science extends far beyond the confines of the tech industry. Every domain, from healthcare to finance, and from agriculture to entertainment, stands to gain from extracting insights from data. However, the inherent technical complexity has often sidelined domain experts who lack formal programming or data analysis training.

As this paper has shown, tools like ChatGPT are bridging this gap. By offering a natural language interface, ChatGPT facilitates a more intuitive interaction with data, enabling professionals from diverse backgrounds to tap into the power of data analysis without first becoming coding experts. A financial analyst, for instance, can ask about trends in stock prices, while a physician can query medical data for patterns.

This democratization has several profound implications:
\begin{enumerate}
  \item \textbf{Enhanced Decision Making:}  Domain experts often possess a nuanced understanding of their fields. Empowering them with data-driven insights can lead to better-informed decisions that factor in both domain knowledge and empirical evidence.

  \item \textbf{Inclusive Innovation:} When more people can harness the power of data, it broadens the scope for innovation. Different perspectives, when backed by data, can lead to breakthroughs that might otherwise remain unexplored.

  \item \textbf{Collaborative Intersections:}  By leveling the playing field, tools like ChatGPT pave the way for interdisciplinary collaborations. Data scientists and domain experts can work in tandem, with each bringing their strengths to the table.
\end{enumerate}

In conclusion, by making data science more accessible, ChatGPT and similar tools are not just democratizing the field; they are catalyzing a movement where insights from data become an inclusive force, driving progress across all sectors.

\subsection{Expanding the Data Science Community and Promoting Diversity}
The democratization of data science doesn't just hold promise for expanded technological and industry advancements; it also carries the potential to diversify the data science community itself. As tools like ChatGPT lower the barriers to entry, individuals from varied backgrounds, cultures, and experiences have the opportunity to engage with and contribute to the field.

As showcased in earlier segments of this paper, the application of AI-driven tools like ChatGPT can empower individuals who may have previously felt excluded due to the high technical demands of traditional data science. Here are a few key takeaways:
\begin{enumerate}
  \item \textbf{Bridging the Educational Gap:}  Not everyone has access to advanced data science training or can afford bootcamps and courses. ChatGPT offers an avenue for these individuals to begin their journey into data analysis, irrespective of their educational background.

  \item \textbf{Promoting Cultural and Gender Diversity:} Historically, STEM fields, including data science, have seen underrepresentation from certain gender and ethnic groups. Democratizing tools can play a part in changing this narrative by making the entry point less intimidating and more accessible.

  \item \textbf{Incorporating Varied Perspectives:}   Data science problems often benefit from diverse viewpoints. As the community grows and diversifies, the variety of perspectives can lead to richer, more comprehensive solutions to complex issues.

  \item \textbf{Fostering Global Participation:}   Geographical barriers are minimized as tools like ChatGPT can be accessed from anywhere. This global accessibility ensures that talent from around the world can contribute to and benefit from data science advancements.
\end{enumerate}

The transformative potential of AI tools like ChatGPT goes beyond technological advancements. By democratizing data science, they pave the way for a more varied and vibrant community of thinkers. This heightened inclusivity enhances the discipline's capacity to address global issues through diverse and comprehensive strategies.

\section{Limitations and Challenges}
\subsection{Introduction}
While the previous sections have painted a promising picture of ChatGPT's potential in the democratization of data science, it's essential to address the inherent challenges and limitations that come with such AI-powered tools. This section will explore the inevitability of biases in training data and the risks associated with the propagation of these biases in AI outputs. We'll also touch upon the limitations in ChatGPT's logical reasoning and the dangers of over-relying on its outputs without human validation. Lastly, we'll discuss the broader ethical implications of AI-facilitated data analysis and the crucial role of human oversight in ensuring responsible AI usage. With these areas in mind, let's delve deeper into the nuances of each concern.

\subsection{Bias and Problematic Outputs}
The training data that powers models like ChatGPT is a reflection of the vast swathes of internet text it has consumed. However, as with any machine learning model, if the data ingested contains biases, the model can also perpetuate and sometimes amplify those biases. In essence, models like ChatGPT could inadvertently become mirrors of the imperfections in our digital discourse.

Consider, for instance, the representation of minority groups, gender biases, or cultural misconceptions present in many online texts. When ChatGPT is trained on such data, there is an inherent risk that the model might generate outputs that reinforce those very biases or misconceptions. Furthermore, when applied to data analysis, these biases could skew insights and lead to misinformed decisions.

While efforts are continually made to refine and retrain models to reduce these biases, the sheer volume and nature of data make it challenging to eliminate them entirely. For data scientists and analysts, it's imperative to be aware of these potential pitfalls and approach AI-generated insights with a degree of skepticism and critical thinking. This underlines the need for a symbiotic relationship between human judgment and AI outputs, ensuring that the insights derived are both accurate and ethically sound.

In addition to biases, another challenge with models like ChatGPT is the potential for incorrect or nonsensical outputs. Given the vast diversity of training data and the model's generative nature, there will inevitably be instances where the generated content might not align with factual accuracy or context. It's crucial for users to recognize that while ChatGPT is a powerful tool, it is not infallible. Misunderstandings or inaccuracies in output can arise from nuances in user queries or the model's interpretation of the data. Users should validate key insights, especially in critical applications, through additional sources or domain-specific expertise. By cross-referencing with established knowledge and understanding the model's limitations, we can harness the benefits of ChatGPT while minimizing the risks of misinformation.

\subsection{Reasoning Limitations of ChatGPT}
ChatGPT, while revolutionary, does not possess human-like reasoning or intuitive capabilities. It operates based on patterns in the data it was trained on. Consequently, while it can generate coherent and often informative responses, it doesn't "understand" them in the way humans do. It lacks a genuine comprehension of context beyond what is present in the query and can't critically evaluate or think abstractly about concepts.

This limitation becomes particularly evident in complex or nuanced situations. For instance, while ChatGPT can perform mathematical calculations or retrieve facts within its training data, it might struggle with hypotheticals that require layered, abstract thinking. Additionally, because its responses are generated based on patterns in the data, it can occasionally produce outputs that, while syntactically correct, may be semantically nonsensical or lack depth.

Furthermore, ChatGPT can't draw from real-time events or updates post its last training cut-off, meaning its knowledge is finite and may not include the very latest developments or trends.

For users of ChatGPT in data science, this means while it's a valuable tool for quick insights, data generation, or answering routine queries, it shouldn't replace human judgment, especially when diving into complex analyses or making significant decisions based on the data.

\subsection{Overreliance and the Importance of Human Oversight}
The advent of tools like ChatGPT makes data analysis more accessible, but with this ease comes the risk of overreliance. An uncritical dependence on ChatGPT might lead users to accept its outputs without questioning or verifying them. Even if the tool can generate technically accurate responses, it doesn't always guarantee they are contextually or logically apt for a specific scenario.

Moreover, while AI can handle vast amounts of data quickly, it can't inherently grasp the nuanced context or the "why" behind certain data patterns. Overreliance on automated systems may also stifle creativity and critical thinking, key elements in data science that drive innovative solutions.

It's also worth noting that every AI system, ChatGPT included, has its blind spots. Whether it's due to biases in training data, the way algorithms were structured, or simply the limitations inherent in a machine's capacity to "understand", these blind spots can lead to oversights or errors.

Therefore, human oversight is imperative. Users should approach ChatGPT as a complementary tool rather than a replacement for human expertise. Humans can provide context, interpret results with skepticism, and ensure that ethical considerations are met. This collaboration between human expertise and AI capabilities can harness the best of both worlds, resulting in richer insights and more informed decisions.

\subsection{Ethical Considerations in AI-powered Data Analysis}
The use of powerful AI tools like ChatGPT for data analysis raises important ethical concerns that the data science community must address.

One key issue is transparency. The "black box" nature of many AI algorithms makes it hard to understand how they arrive at results. Without transparency, users may struggle to trust or validate the outputs. This creates ethical dilemmas when these opaque results inform real-world decisions.

In addition, biases in training data can lead AI systems to perpetuate harmful stereotypes and generate skewed or unfair results that disadvantage certain groups. For example, biased telematics data could produce discriminatory insurance premiums.

Privacy is another vital ethical consideration. When analyzing large datasets, there is always a risk of inadvertently revealing sensitive information or identifying individuals, especially when merging or analyzing high-dimensional data.

Finally, deploying AI systems raises questions of accountability. If something goes wrong, who is responsible - the developers, the users, or the institutions deploying the technology?

To navigate these challenges responsibly, the data science field needs a multidisciplinary approach that draws from computer science, ethics, sociology, and other domains. As adoption of AI tools like ChatGPT increases, prioritizing ethical considerations will be key to ensuring fair and beneficial outcomes. The industry must proactively address transparency, bias, privacy, accountability, and other ethical issues.

\section{Conclusion}

As the landscape of data science evolves, tools like ChatGPT offer transformative opportunities. They not only streamline intricate processes but also democratize access, inviting diverse minds into the realm of data exploration. Throughout this paper, we've unveiled the robust capabilities of ChatGPT, from data cleaning to sophisticated analysis.

Nevertheless, the capabilities of tools like ChatGPT come with their own complexities and nuances. While AI models like ChatGPT are exceptionally powerful, relying on them without a thorough understanding can lead to unintended outcomes and misconceptions.

But, viewed through a lens of optimism, these challenges also represent an invitation—to learn, to question, and to collaborate. As we stand on the threshold of a new era of data science, one fueled by the symbiosis of humans and AI, our collective responsibility is to tread wisely. By doing so, we can harness the promise of AI-facilitated data analysis, shaping a future where data-driven insights are both profound and accessible to all.

 \end{document}